\documentclass[aps,prb,twocolumn,shortbibliography,superscriptaddress]{revtex4-1}
\usepackage{epsfig}
\usepackage{epstopdf}
\usepackage{amsmath}
\usepackage{amsfonts}
\usepackage{amssymb}
\usepackage{hyperref}
\usepackage{bm}
\usepackage{makecell}
\usepackage{rotating}
\usepackage{hyperref}

\usepackage{graphicx}
\usepackage{dcolumn}
\usepackage{bm}
\usepackage{color}

\usepackage{tikz,xcolor,hyperref}

\definecolor{lime}{HTML}{A6CE39}
\DeclareRobustCommand{\orcidicon}{%
	\begin{tikzpicture}
	\draw[lime, fill=lime] (0,0)
	circle [radius=0.16]
	node[white] {{\fontfamily{qag}\selectfont \tiny ID}};
	\draw[white, fill=white] (-0.0625,0.095)
	circle [radius=0.007];
	\end{tikzpicture}
	\hspace{-2mm}
}

\foreach \x in {A, ..., Z}{%
	\expandafter\xdef\csname orcid\x\endcsname{\noexpand\href{https://orcid.org/\csname orcidauthor\x\endcsname}{\noexpand\orcidicon}}
}

\begin{document}

\title{Dirac surface states, multiorbital dimerization and superconductivity \\ in Nb- and Ta-based A15 compounds}

\author{Raghottam M. Sattigeri\orcidE}
\email{rsattigeri@magtop.ifpan.edu.pl}
\affiliation{International Research Centre Magtop, Institute of Physics, Polish Academy of Sciences, Aleja Lotnik\'ow 32/46, PL-02668 Warsaw, Poland}

\author{Giuseppe Cuono\orcidA}
\email{gcuono@magtop.ifpan.edu.pl}
\affiliation{International Research Centre Magtop, Institute of Physics, Polish Academy of Sciences,
Aleja Lotnik\'ow 32/46, PL-02668 Warsaw, Poland}

\author{Ghulam Hussain\orcidB}
\affiliation{International Research Centre Magtop, Institute of Physics, Polish Academy of Sciences,
Aleja Lotnik\'ow 32/46, PL-02668 Warsaw, Poland}

\author{Xing Ming}
\affiliation{College of Science, Guilin University of Technology, Guilin 541004, People’s Republic of China.}

\author{Angelo Di Bernardo\orcidF}
\affiliation{Dipartimento di Fisica ’E.R. Caianiello’, Universit\'a degli Studi di Salerno,
via Giovanni Paolo II 132, I-84084 Fisciano (SA), Italy}

\author{Carmine Attanasio\orcidG}
\affiliation{Dipartimento di Fisica ’E.R. Caianiello’, Universit\'a degli Studi di Salerno,
via Giovanni Paolo II 132, I-84084 Fisciano (SA), Italy}

\author{Mario Cuoco\orcidC}
\email{mario.cuoco@spin.cnr.it}
\affiliation{Consiglio Nazionale delle Ricerche CNR-SPIN, UOS Salerno, I-84084 Fisciano (Salerno),
Italy}
\affiliation{Dipartimento di Fisica ’E.R. Caianiello’, Universit\'a degli Studi di Salerno,
via Giovanni Paolo II 132, I-84084 Fisciano (SA), Italy}

\author{Carmine Autieri\orcidD}
\email{autieri@magtop.ifpan.edu.pl}
\affiliation{International Research Centre Magtop, Institute of Physics, Polish Academy of Sciences,
Aleja Lotnik\'ow 32/46, PL-02668 Warsaw, Poland}

\date{\today}
\begin{abstract}
Using first-principle calculations, we investigate the electronic, topological and superconducting properties of Nb$_3$X (X = Ge, Sn, Sb) and Ta$_3$Y (Y = As, Sb, Bi) A15 compounds. We demonstrate that these compounds host Dirac surface states which are related to a nontrivial $\mathbb{Z}_2$ topological value. The spin-orbit coupling (SOC) splits the highly degenerate R point close to the Fermi level enhancing the amplitude of the spin Hall conductance. Indeed, despite the moderate spin-orbit of the Nb-compounds, a large spin Hall effect is also obtained in Nb$_3$Ge and Nb$_3$Sn compounds. We show that the Coulomb interaction opens the gap at the R point thus making the occurrence of Dirac surface states more obvious. We then investigate the superconducting properties by determining the strength of the electron-phonon BCS coupling. The evolution of the critical temperature is tracked down to the 2D limit indicating a reduction of the transition temperature which mainly arises from the suppression of the density of states at the Fermi level.
Finally, we propose a minimal tight-binding model based on 
three coupled Su-Schrieffer-Heeger chains with t$_{2g}$ Ta- and Nb-orbitals reproducing the spin-orbit splittings at the R point among the $\pi$-bond bands in this class of compounds. We separate the kinetic parameters in $\pi$ and $\delta$-bonds, in intradimer and interdimer hoppings and discuss their relevance for the topological electronic structure.
We point out that Nb$_3$Ge might represent a $\mathbb{Z}_2$ topological metal with the highest superconducting temperature ever recorded.
\end{abstract}


\maketitle

\section{Introduction}

Topological superconductivity is a captivating phase of condensed matter physics characterized by unique electronic properties such as the Majorana fermions\cite{Majorana}. These particles, distinct for their non-Abelian statistics, hold promise for fault-tolerant quantum computation. This enigmatic phase of matter has ignited a surge of research, with potential applications spanning quantum computing to quantum information storage. Understanding and harnessing topological superconductivity holds the key to obtaining new quantum technologies. Therefore, a growing interest in topological superconductivity has been rising in the last decade. \cite{Yonezawa2017,Sharma_2022}
Topological superconductivity can arise from the coexistence of Bardeen–Cooper–Schrieffer (BCS) superconductivity and Dirac states which can lead to mixed pairing order parameters or topological superconductivity, therefore, the Dirac surface states of superconductors are platforms for investigating the interplay between superconductivity and topologically nontrivial Fermi surfaces\cite{PhysRevLett.113.046401,Gao2022}.

The most common k-space topological phase is characterized by a non-zero $\mathbb{Z}_2$ topological invariant. The $\mathbb{Z}_2$ topological insulators have gapped bulk band structure and gapless surface states. These surface states are protected by time-reversal symmetry.
$\mathbb{Z}_2$ topological metals are conducting materials with gapless bulk band structures and gapless surface states\cite{Cheng2023,PhysRevLett.116.016401,PhysRevB.106.064109}.
In the last years, different families of materials have been proposed to be $\mathbb{Z}_2$ topological metals with a superconductive ground state. We can mention some of the most representative members of these families as the kagome\cite{PhysRevLett.125.247002} CsV$_3$Sb$_5$, the non-symmorphic ZrOSSi\cite{Toni2023}, beryllenes\cite{LI2023101257} and KHgAs compounds\cite{Dai2023} and the van der Waals Ta$_2$Pd$_3$Te$_5$ material\cite{Higashihara2021,PhysRevB.107.L020503}. Dirac surface states were also found in several undoped iron-based systems such as BaFe$_2$As$_2$ and LiFeAs\cite{Zhang2018,PhysRevB.106.115114}.

The Nb-based A15 compounds were widely studied in the past due to their superconductivity with a high critical temperature ($T_c$).
The superconductivity in Nb-based A15 compounds was found to be BCS-like \cite{Stewart15}, namely the pairing of the superconducting electrons is via electron–phonon coupling.
The Fermi level of these compounds is close to a peak in the density of states deriving from dimerized one-dimensional Nb chains.
In silicides and germanides of transition metals, the highest $T_c$ was found in V$_3$Si among all the known binary compounds \cite{Hardy54}.
The A15 claimed the title of the highest $T_c$ superconductors in 1954 when $T_c$ = 18 K was first observed in Nb$_3$Sn\cite{Matthias54}.
Additionally, high-temperature superconductivity in H-based A15 compounds was recently found \cite{cross2023hightemperature} 
Other Nb-based superconductors were then found as for example Nb$_3$Al  with $T_c$=18.8 K\cite{Willens69}, Nb$_3$Ga with $T_c$=20.3 K in\cite{Webb71} and Nb$_3$Ge with $T_c$=22.3 K\cite{Gavaler03}. Recently, several Nb-based compounds were investigated for their exotic superconducting properties such as the van der Waals NbX$_2$(X=S, Se)\cite{PhysRevB.105.035119,PhysRevB.102.155115,Cossu2020} and the non-centrosymmetric NbRe\cite{10.1063/1.4997675,MakhdoumiKakhaki2023}
Both theoretical and experimental investigations have extensively explored the properties of A15 compounds \cite{Sundareswari10,Muller80,Wu20,Stewart15,Yanlong14,DeMarzi13} based on niobium and tantalum, due to their high critical temperature and high critical magnetic fields.
Due to the discovery of the unconventional high $T_c$ superconductivity in heavy fermions and cuprates, the superconductive phase of the A15 compounds has received less scientific attention in recent years.
Recently, the A15 compounds regained attention due to the large degeneracy at the R point, these A15 compounds are multifold fermion metals\cite{doi:10.1126/science.aaf5037} with a notable spin-Hall effect\cite{derunova2019giant,Hou21} and non-trivial band structure topology\cite{Kim19,PhysRevLett.116.186402}. Dirac points are emergent along the R–M path due to the C$_4$ rotational symmetry.\cite{Hou21} The large spin-Hall conductivity in these compounds is due to the fact that they have bands close to the Fermi level that present crossings unprotected under the action of the spin-orbit coupling interaction (SOC) \cite{derunova2019giant}. The Ta-based A15 compounds have the same filling as the Nb-based A15 compounds, with the Ta having a larger SOC. The Ta$_3$Sb compound with A15 crystal structure was proposed to be a topological superconductor\cite{PhysRevB.105.184510,PhysRevB.107.104504}.

Regarding the realization of devices, Nb$_3$Sn superconductors have significant applications in constructing high-field magnets\cite{Xu_2017}.
Nb$_3$Sn can be used as a coating for producing superconducting
surfaces\cite{10.1063/5.0015376} and for particle accelerators\cite{Sundahl2021}. 
Nb$_3$Sn thin films are promising candidates for future applications in superconducting radio frequency cavities\cite{10.1063/5.0015376}.

\begin{figure}[t!]
\centering
\includegraphics[width=0.85\linewidth]{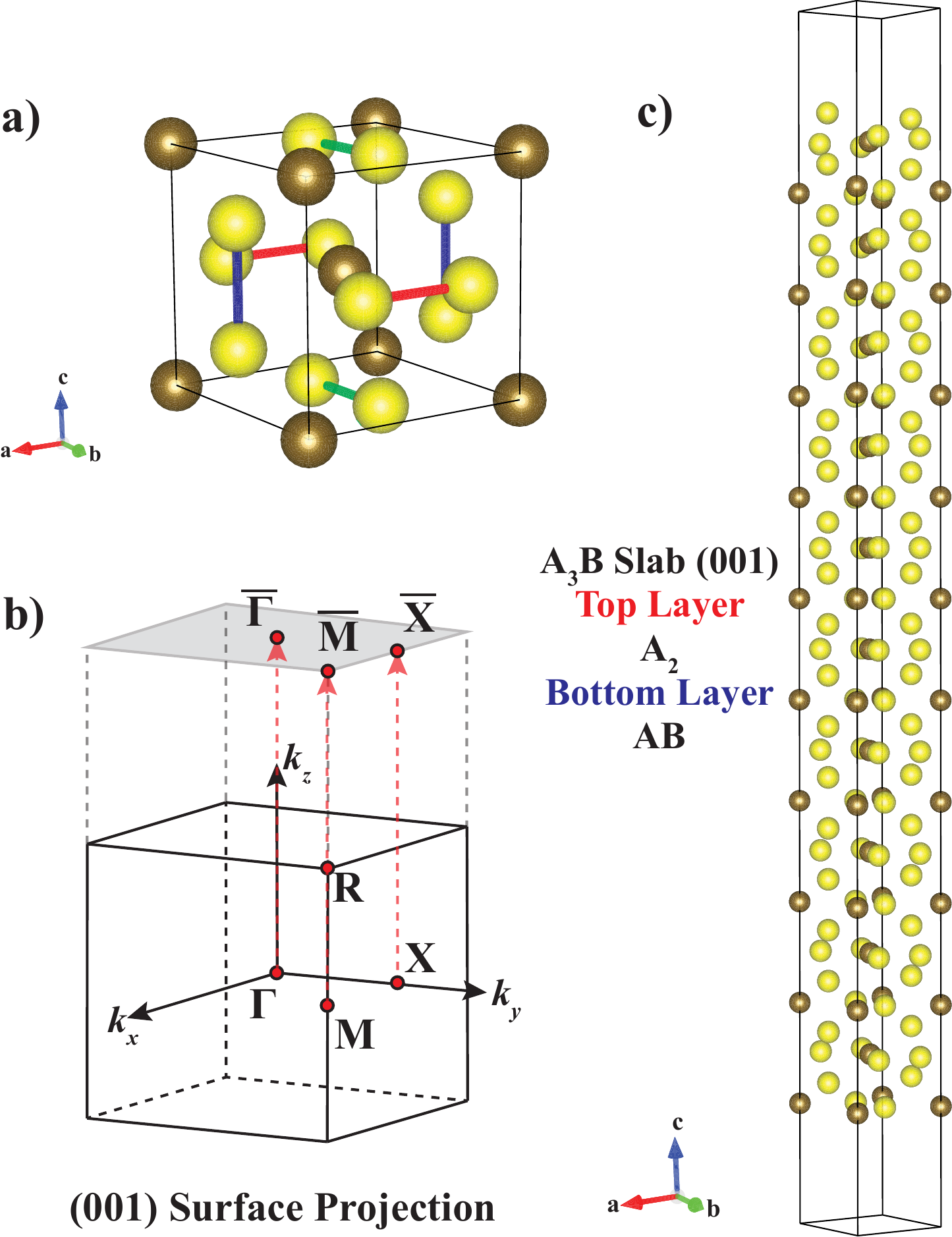}
\caption{(a) Crystal structure of the A$_3$B compound. The $A$ sites are shown in yellow and $B$ sites are shown in brown. The A sites belong to the dimers along the $a$, $b$ and $c$-axis of the unit cell. We indicate with a red bond the dimer along the a-axis, with a green bond the dimer along the b-axis and with a blue bond the dimer along the c-axis. (b) The irreducible bulk Brillouin zone with (001) surface projections is used for computing surface states. (c) Slab of A$_3$B compounds periodic along (001) crystal direction with the top surface layer composed of A$_2$ and bottom surface layer composed of AB elements. The composition of the bottom and top surface layers is preserved along the paper.}
\label{Crystal structure}
\end{figure}

In this paper, we study the electronic, topological and superconductive properties of Nb$_3$X (X = Ge, Sn, Sb) and Ta$_3$Y (Y = As, Sb, Bi) A15 compounds and we demonstrate that all these compounds are $\mathbb{Z}_2$ topological metals hosting Dirac surface states. We study the interplay between the electronic and topological properties with the spin-Hall and BCS superconductivity. These topological properties can be explained by a tight-binding model with three coupled  Su–Schrieffer–Heeger (SSH) chains. Nb$_3$Sb and Ta$_3$Y (Y = As, Sb, Bi) have half-filling p- and d-orbitals, while Nb$_3$Ge and Nb$_3$Sn have one electron less. The paper is organized as follows.
In the next Section, the results of our {\it ab initio} calculations are reported. In more detail, this Section is divided into many Subsections: in Subsection A the structural and electronic properties of Nb$_3$X and Ta$_3$Y are investigated, in Subsection B we study the Spin Hall conductivity, while in Subsections C and D we discuss the topological properties for the Ta-based and Nb-based compounds, respectively. Subsection E is devoted to the superconductivity, while Subsection G is dedicated to the thickness-dependent density of states. In Section III, we report our tight-binding model composed of the three coupled chains of the SSH model with t$_{2g}$ orbital basis. Finally, Section IV is devoted to the discussion, conclusions and outlook.

\section{Results}

\subsection{Structural and Electronic properties of Nb$_3$X and Ta$_3$Y}

\begin{figure*}[t!]
\centering
\includegraphics[width=\linewidth]{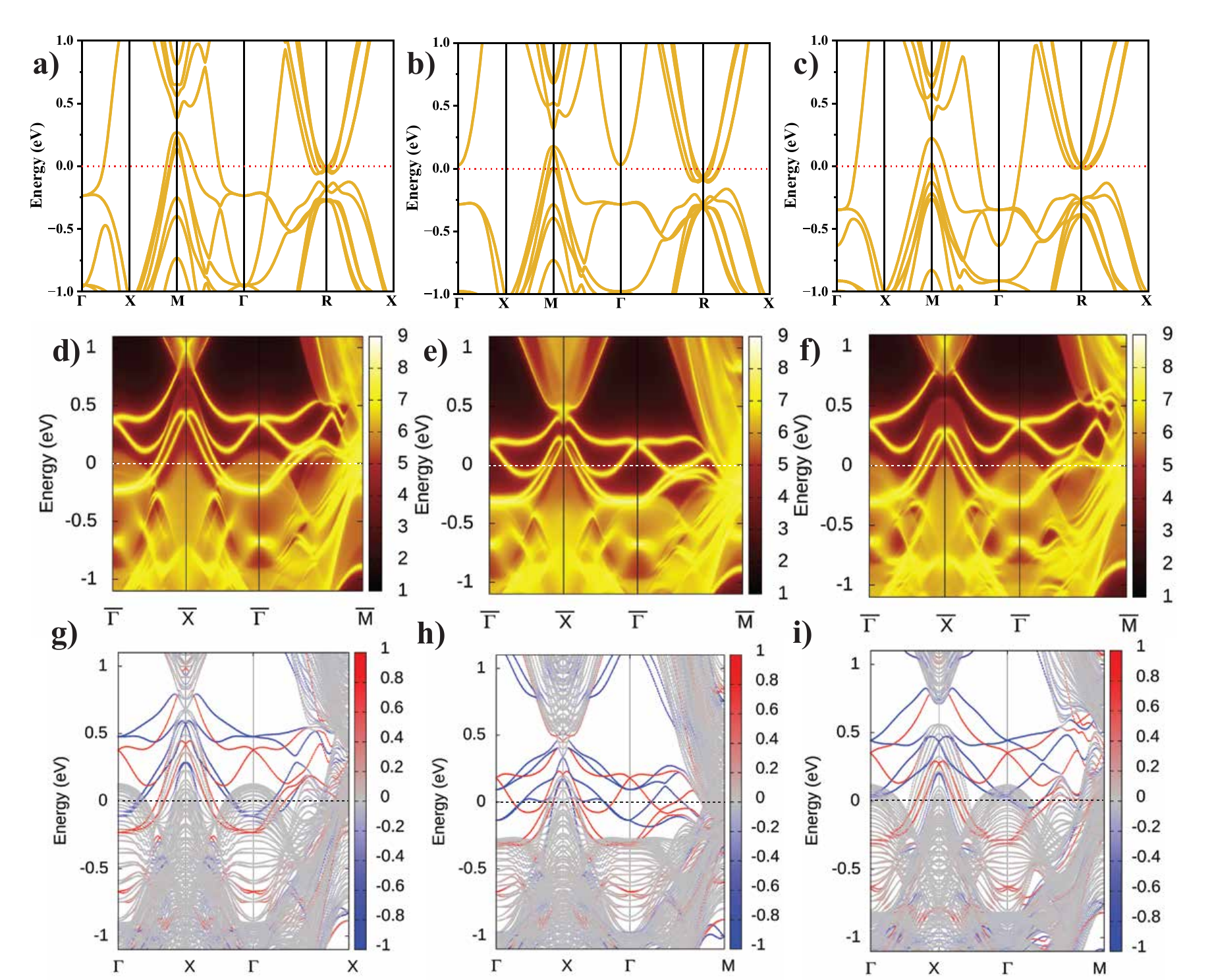}
\caption{Electronic structures of (a) Ta$_3$As, (b) Ta$_3$Sb and (c) Ta$_3$Bi with spin-orbit coupling. Surface states of (d) Ta$_3$As, (e) Ta$_3$Sb and (f) Ta$_3$Bi indicating spin-momentum locked Dirac dispersions at $\Gamma$ point. Slab band structures of (g) Ta$_3$As, (h) Ta$_3$Sb and (i) Ta$_3$Bi with the red bands indicating the contribution from the top surface layer and the blue bands indicating the contribution from the bottom surface layer of the slab presented in Fig \ref{Crystal structure}. The legends on the right in panels (d-i) indicate the spectral weights of the top and bottom layers. The Fermi level is set to zero in all panels.}
\label{bands-ta3x}
\end{figure*}

A15 compounds are governed by the \textit{Pm$\overline{3}$n} (No. 223) space group which exhibits an intermetallic nature arising from a chemical composition of A$_3$B, where site A is occupied by a transition metal/d-block element and site B is occupied by the p-block element. The crystal structure presented in Fig. \ref{Crystal structure}(a) is a typical unit cell of an A15 compound with inversion symmetry containing eight atoms with site A forming one-dimensional chains along the edges which are orthogonal to neighboring faces and the B site forming a body-centered cubic lattice. The presence of spatial inversion symmetry is due to the non-symmorphic space group governing the system which involves a screw axis in the [001] crystal direction. The A sites in A$_3$B composition occupy 6c Wyckoff positions (0.25,0.00,0.50) and the B sites occupy 2a Wyckoff positions (0.00,0.00,0.00). The crystal structure presents three dimers of the A atoms, along the $a$, $b$ and $c$ axes, as shown in Fig. \ref{Crystal structure}(a). A typical slab of A15 compound periodic in [001] crystal direction is presented in Fig. \ref{Crystal structure}(c) with surface Brillouin zone highlighted in Fig. \ref{Crystal structure}(b). Such slab structure leads to the absence of the fractional translation symmetries which breaks the four-fold symmetry in bulk to two-fold symmetry on the surfaces.\cite{Kim19} Therefore the surface Brillouin zone would be orthorhombic, such surface projection has the momentum path $\Gamma \rightarrow$ X $\rightarrow \Gamma \rightarrow$ M/S (since on the surface the S and M points of the orthorhombic and cubic Brillouin zone are equivalent).\cite{Kim19,derunova2019giant} This exposes two unique surfaces, the top surface originates from the A$_2$ atomic arrangement and the bottom surface originates from the AB atomic arrangement. The optimized lattice constants (a) after structural relaxation for Nb$_3$Ge, Nb$_3$Sn and Nb$_3$Sb are 5.177 {\AA}, 5.324 {\AA} and 5.303 {\AA}, respectively which are in agreement with literature.\cite{devantay1981physical, tarutani1977atomic,furuseth1964arsenides} While the optimized lattice constants (a) for Ta$_3$As, Ta$_3$Sb and Ta$_3$Sn are, 5.203 {\AA}, 5.329 {\AA} and 5.394 {\AA}, respectively.

\begin{figure*}[t!]
\centering
\includegraphics[width=\linewidth]{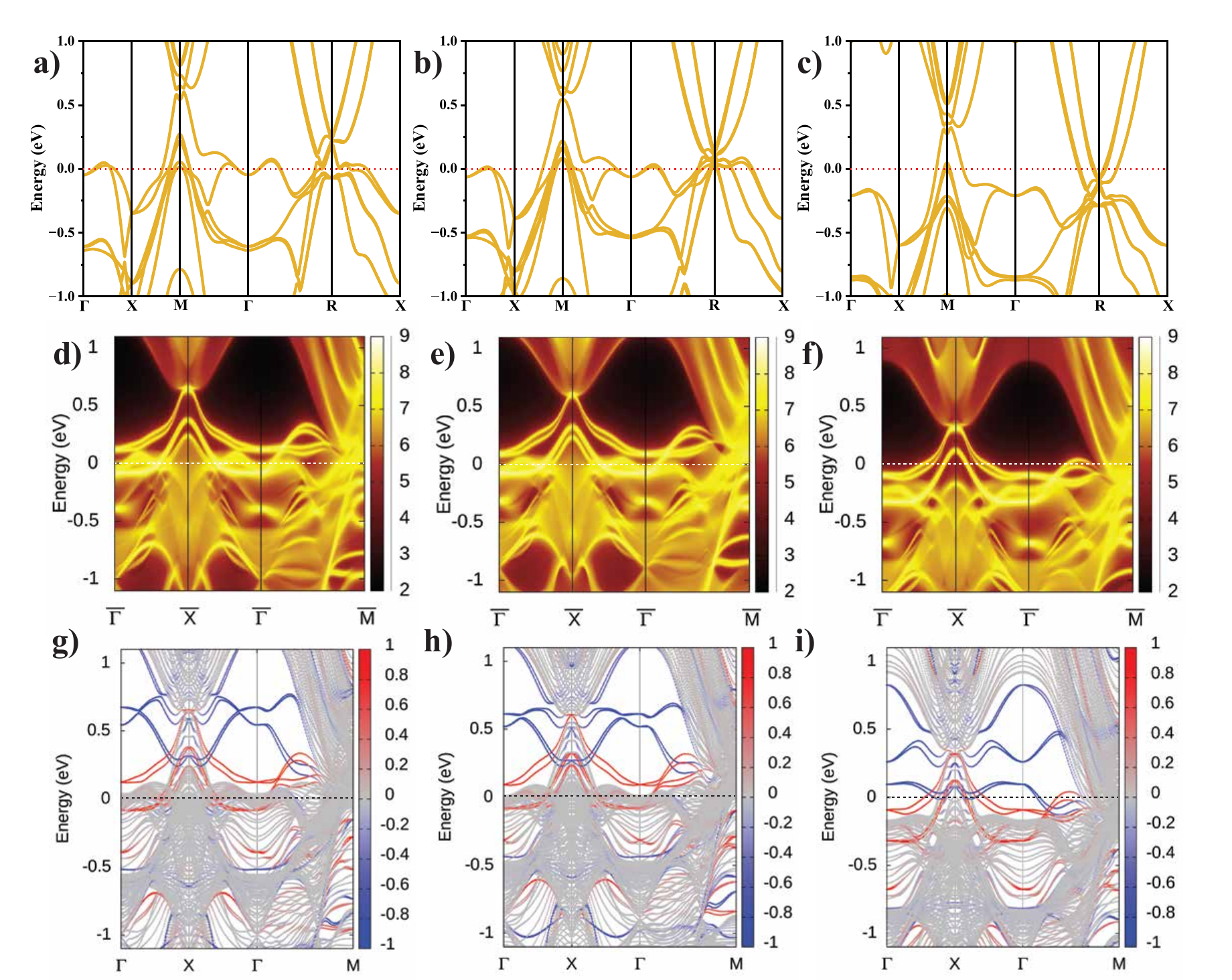}
\caption{Electronic structures of (a) Nb$_3$Ge, (b) Nb$_3$Sn and (c) Nb$_3$Sb with spin-orbit coupling. Surface states of (d) Nb$_3$Ge, (e) Nb$_3$Sn and (f) Nb$_3$Sb indicating spin-momentum locked Dirac dispersions at $\Gamma$ point. Slab band structures of (g) Nb$_3$Ge, (h) Nb$_3$Sn and (i) Nb$_3$Sb with the red bands indicating the contribution from the top surface layer and the blue bands indicating the contribution from the bottom surface layer of the slab presented in Fig \ref{Crystal structure}. The legends on the right in panels (d-i) indicate the spectral weights of the bottom and top layers. The Fermi level is set to zero in all panels.}
\label{bands-nb3y}
\end{figure*}

The computational framework is described in Appendix B.
In Figs. \ref{bands-ta3x} and \ref{bands-nb3y}  the electronic structures of Ta$_3$Y and Nb$_3$X compounds are shown, respectively. Both band structures host fourfold rotational symmetries. This is due to the presence of non-symmorphic symmetry operations involving fourfold rotations and fractional translations with respect to the [001] crystal directions. Apart from the symmetries due to time-reversal and spatial inversion symmetry, we observe additional degeneracies in the momentum. Due to the non-symmorphic symmetries and screw axes, we have a 4-fold degeneracy above Fermi which splits into 8-fold degeneracy due to SOC whereas, the 6-fold degeneracy below the Fermi level splits into 4-fold and 8-fold degeneracies with the 4-fold degeneracies lying above the 8-fold degenerate states at the R point due to SOC. However, in the case of Ta$_3$Bi due to strong SOC originating from Bi atoms, the 6-fold degenerate states below Fermi splits similar to other compounds but with the exception that the 4-fold degeneracies are pushed below the 8-fold degeneracies. Such 8-fold degenerate points are known to represent double Dirac points.\cite{PhysRevLett.116.186402}
As we demonstrate in Appendix A, in the low-energy sector at the R-point two groups of bands are present: the bands related to the $\pi$ and $\delta$ intradimer bonds. In Figs. \ref{bands-ta3x}(a,b,c) and  \ref{bands-nb3y}(a,b,c) the band structures in the range between -1 eV and 1 eV are shown, at the R point we can see in this range the $\pi$-bond bands. The $\delta$-bond bands are around 1.5 eV below the Fermi level. The parabolic band appearing at the $\Gamma$ point for the Ta-based compounds is the 6s band of Ta. The 6s band crosses the p-d bands for Ta$_3$As and Ta$_3$Sb and increases the density of states.

However, at the R point in the momentum space, we have mild variations in Ta$_3$Y and Nb$_3$X compounds with eigenvalues that are four and eight times degenerate as shown in Figs. \ref{bands-ta3x} and \ref{bands-nb3y} which is in agreement with the literature. While in Ta-based compounds, the strong SOC at the R-point opens the gap between $\pi$-orbital bands, as we can see in Fig. \ref{bands-ta3x} (a, b, c), in the Nb-based there is a smaller splitting at R but the bands with the camelback shape around R are still crossing keeping conduction and valence sectors entangled, as it is shown in Figs. \ref{bands-nb3y} (a, b,c). 

The crystal symmetries are responsible for orbital hybridizations in the compounds, once we apply the SOC the anticrossings strongly contribute to the Berry curvature and spin Berry curvature. The multiple crossings observed in the electronic structures here give rise to large spin Berry curvatures which effectively translates to a large spin Hall effect in such compounds since the Fermi level lies within gaped crossings. 

At the Fermi level, we observe a high density of states with large contributions (due to multiple crossings) from transition metals Ta or Nb and minor contributions from other constituent elements of the A$_3$B composition, as shown in Appendix C. In the momentum space, the conduction band minima at $\Gamma$ point originates from the s-orbitals of group III elements or pnictogens in the A$_3$B composition while the valence band maxima at $\Gamma$ point originate from the d-orbitals of transition metals. Since these dispersions vary between the Ta$_3$Y and Nb$_3$X compositions, their relative positions define the density of states at the Fermi level and in turn the superconducting critical temperature ($T_c$) of the system.

In Ta$_3$As and Ta$_3$Bi, the conduction band minima with s-orbital contributions are below the valence band maxima with d-orbital contributions at $\Gamma$ point as observed in Fig. \ref{bands-ta3x} (a,c).  However, in the case of Ta$_3$Sb and Nb$_3$X, the conduction bands and valence bands are well separated throughout the Brillouin zone as presented in Fig. \ref{bands-ta3x}(b) and Fig. \ref{bands-nb3y}, respectively. In this last case, the s-orbital contributions are farther away from the Fermi level as compared to the Ta$_3$Y family. The well-resolved band manifolds in Ta$_3$Sb and Nb$_3$X are accompanied by band inversions across the Fermi level which could produce topological properties.
We will see that the presence of the s-orbitals band produces additional anticrossings and additional bands that increase the spin Hall conductivity (SHC) and $T_c$ respectively.

\subsection{Spin Hall conductivity}

\begin{figure}[h!]
\centering
\includegraphics[width=\linewidth]{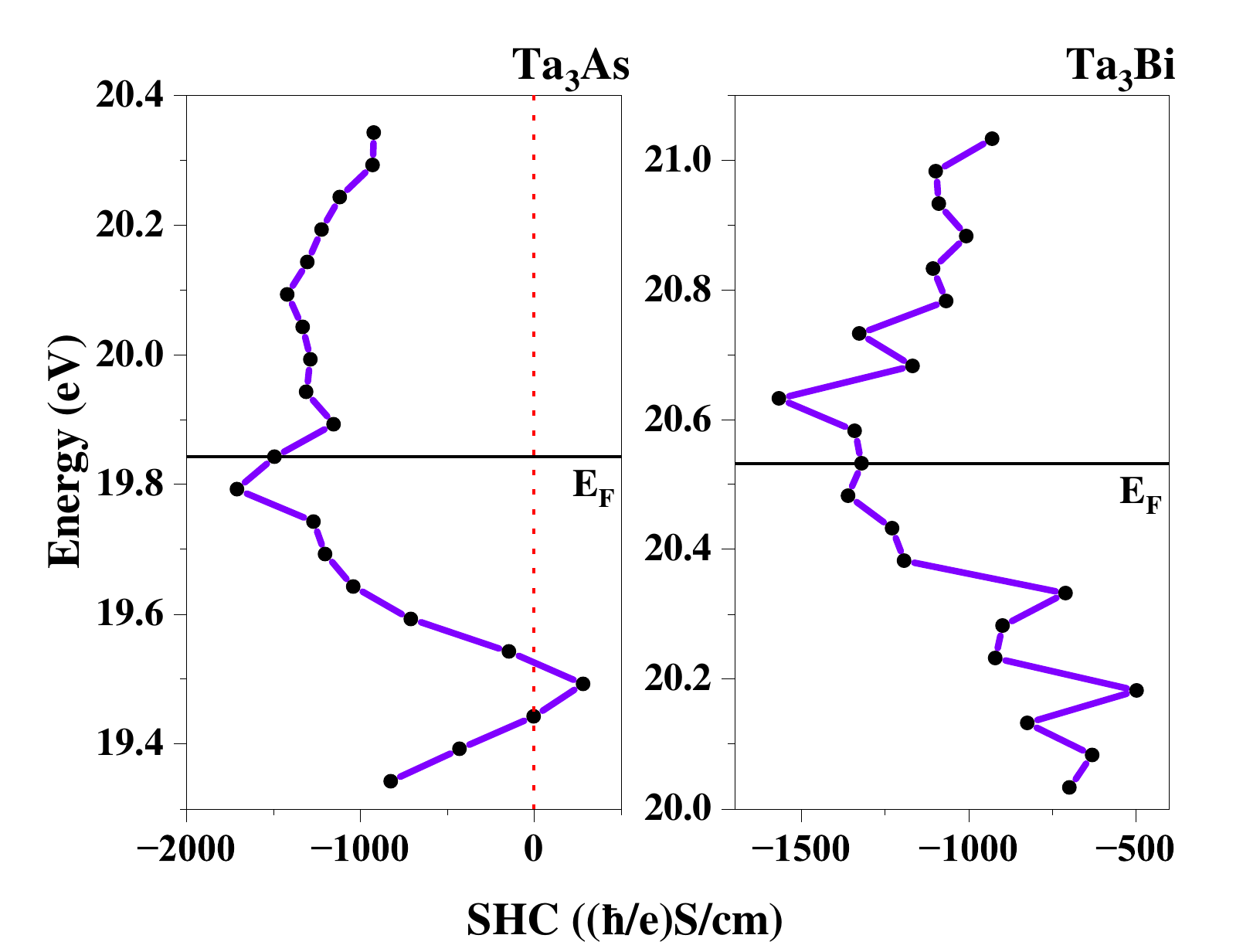}
\caption{Spin Hall conductivity as a function of the energy for Ta$_3$As and Ta$_3$Bi compounds. The black solid line indicates Fermi level.}
\label{ta3x-SHC}
\end{figure}

It is clear from the electronic structures of these compounds that they host multiple crossings and anticrossings, therefore we have a large change in the spin Berry curvature which indicates that the spin Hall effect should be large. It is known that the SHC is inversely proportional to the spin-orbit induced gap. Accordingly, in the case of Ta$_3$As, Ta$_3$Sb and Ta$_3$Bi (spin-orbit induced gap in electronic structure in increasing order) we find that the SHC at the Fermi level is $-$ 1492.8 ($\hbar/e$) Scm$^{-1}$, $-$ 1423.86 ($\hbar/e$) Scm$^{-1}$ and $-$ 1320.2 ($\hbar/e$) Scm$^{-1}$ respectively (which has decreasing trend as compared to the increasing order of the spin-orbit induced gap).
The SHC of Ta$_3$As and Ta$_3$Bi close to the Fermi level are shown in Fig. \ref{ta3x-SHC}.
The SHC exhibits a wide peak that encompasses the energy range where the gapped crossings are located. In both cases of Ta$_3$As and Ta$_3$Bi the peak is very close to the Fermi level since the gapped crossings are located near the Fermi level, as shown in Fig. \ref{bands-ta3x}(a,c).
A similar trend is observed in the case of Nb$_3$X (with group III elements) i.e., for Nb$_3$Ge and Nb$_3$Sn (with the spin-orbit induced gap in electronic structure in increasing order) we have SHC at the Fermi level of $-$ 1691.4 ($\hbar/e$) Scm$^{-1}$ and $-$ 983.1 ($\hbar/e$) Scm$^{-1}$ respectively. In the composition of Nb$_3$X, Nb$_3$Sb is an outlier with significantly low and positive SHC of 155.3 ($\hbar/e$) Scm$^{-1}$ since it is a pnictogen substitution as compared to the other two which are group III elements.

Large values of the SHC are usually associated with robust orbital texture. Indeed, the A15 systems host a robust orbital texture.\cite{Kim19}
Even if the orbital moment is zero, since the compounds present inversion symmetry and highly symmetric crystal structure,
the breaking of the inversion symmetry at the surface will generate an orbital magnetic moment. This makes the study of the surface states of A15 compounds interesting.

\begin{figure*}[ht!]
\centering
\includegraphics[width=\linewidth]{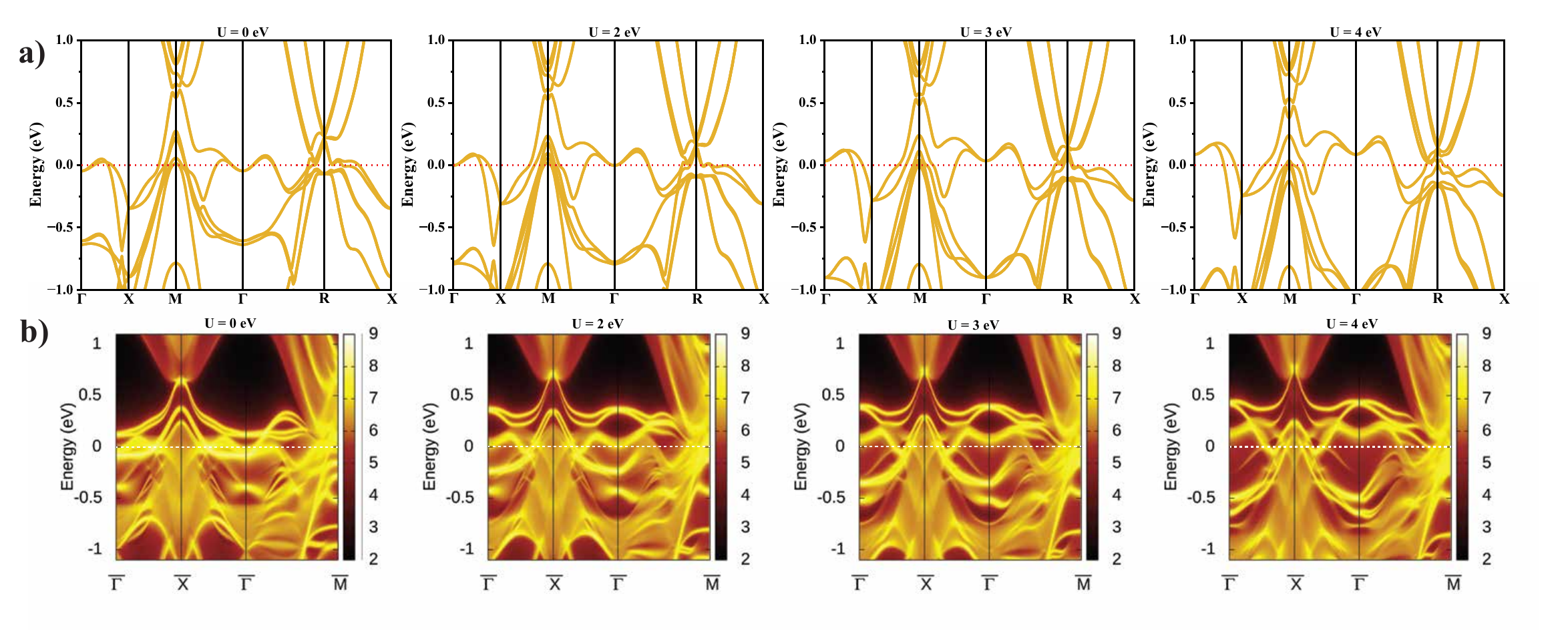}
\caption{(a) Electronic structure of Nb$_3$Ge for U = 0, 2, 3 and 4 eV with spin-orbit coupling. (b) Surface states of Nb$_3$Ge for U = 0, 2, 3 and 4 eV indicating the presence of spin-momentum locked Dirac dispersions shifting farther from the Fermi level with the increase in U. The legends on the right in panel (b) indicate the spectral weights. The Fermi level is set to zero in all panels.}
\label{bands_Nb3Ge}
\end{figure*}

\begin{figure*}[ht!]
\centering
\includegraphics[width=\linewidth]{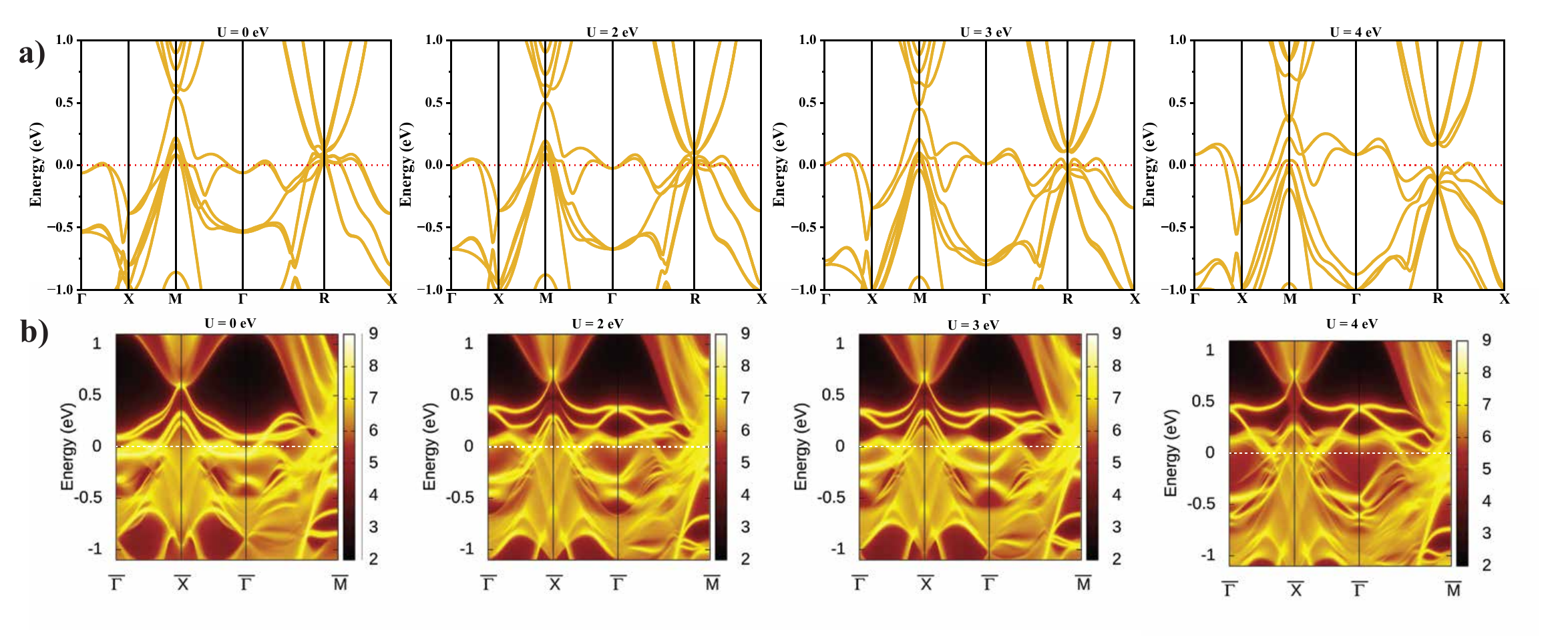}
\caption{(a) Electronic structure of Nb$_3$Sn for U = 0, 2, 3 and 4 eV with spin-orbit coupling. (b) Surface states of Nb$_3$Sn for U = 0, 2, 3 and 4 eV indicating the presence of spin-momentum locked Dirac dispersions shifting farther from the Fermi level with the increase in U. The legends on the right in panel (b) indicate the spectral weights. The Fermi level is set to zero in all panels.}
\label{bands_Nb3Sn}
\end{figure*}

\begin{figure*}[ht!]
\centering
\includegraphics[width=\linewidth]{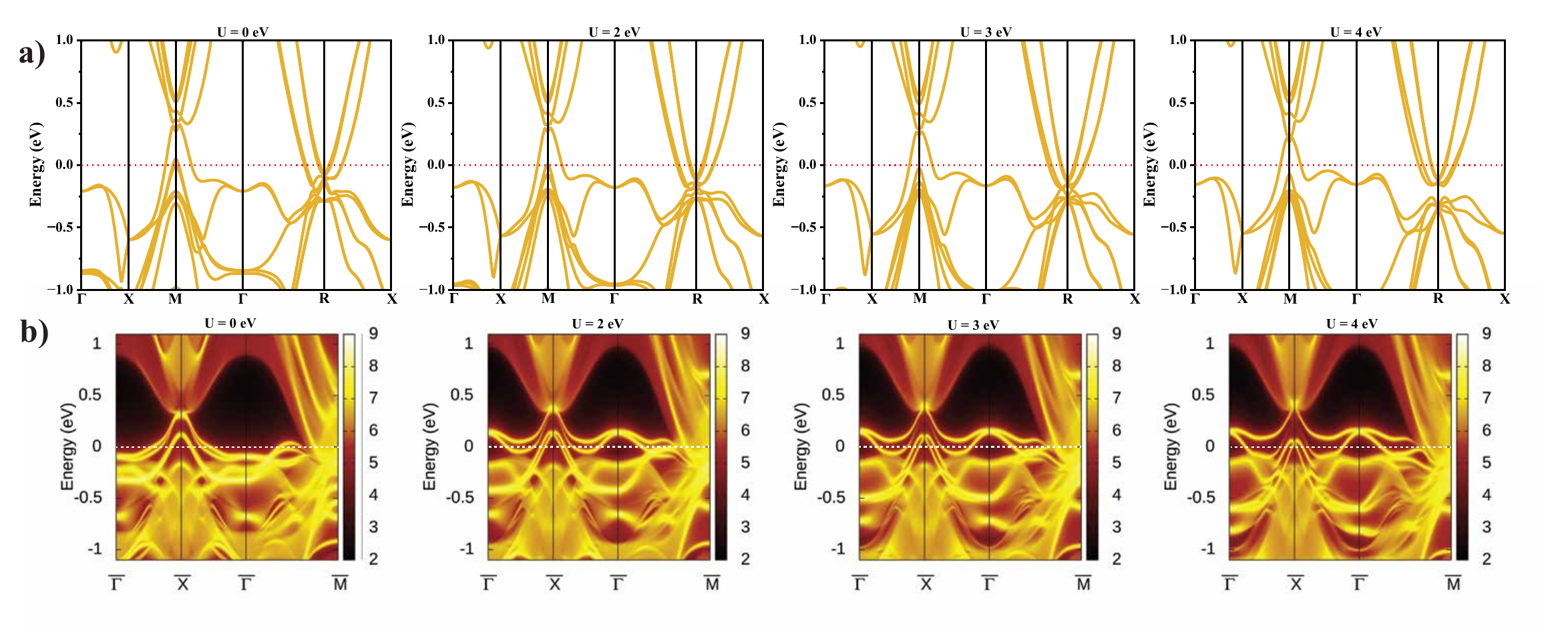}
\caption{(a) Electronic structure of Nb$_3$Sb for U = 0, 2, 3 and 4 eV with spin-orbit coupling. (b) Surface states of Nb$_3$Sb for U = 0, 2, 3 and 4 eV indicating the presence of spin-momentum locked Dirac dispersions shifting farther from the Fermi level with the increase in U. The legends on the right in panel (b) indicate the spectral weights. The Fermi level is set to zero in all panels.}
\label{bands_Nb3Sb}
\end{figure*}

\subsection{Topological properties of Ta$_3$Y (Y = As, Sb, Bi)}

In this subsection, we discuss first the topological properties of the Ta$_3$Sb that is the ideal case, later, we discuss the topological properties of Ta$_3$As and Ta$_3$Bi with the presence of the s-band at the Fermi level.
Since the Ta$_3$Sb compound shows well-resolved band manifolds, we compute the surface states projected on [001] crystal direction as presented in Fig. \ref{bands-ta3x}(e). Clearly, this compound hosts spin-momentum locked surface states with Dirac dispersions at $\Gamma$ point.
We also represent the corresponding slab band structure in Fig. \ref{bands-ta3x}(h), which shows that the Dirac dispersion at $\Gamma$ originates from the top surface layers.
Although these are slightly away from the Fermi level, one can realize them at the Fermi level in experimental conditions by varying the carrier concentrations. Albeit, as the band manifolds are well resolved in the case of Ta$_3$Sb, we compute the $\mathbb{Z}_2$ topological invariants using the Wilson loop method around the Wannier charge centers. Therefore, for Ta$_3$Sb, the four $\mathbb{Z}_2$ 3D topological invariants are  ($\nu_0$,$\nu_1\nu_2\nu_3$)=(1;000) indicating a strong topological insulator character. 

Since the conduction bands and valence bands are degenerate at the $\Gamma$ point due to the presence of the 6s band in the case of Ta$_3$As and Ta$_3$Bi, we do not calculate the $\mathbb{Z}_2$ invariants for these compounds. Although $\mathbb{Z}_2$ is not well-defined, from the band structure we can see the Dirac surface states for Ta$_3$As and Ta$_3$Bi as shown in Figs. \ref{bands-ta3x}(d,g) and \ref{bands-ta3x}(f,i), respectively, where we show both the surface states and the slab band structures. However, when we include the 6s band in the tight-binding model with the Wannier basis, the Dirac surface states are blurred by the hybridization with the 6s band (see Appendix E for more details). The Dirac surface states would be difficult to detect in Ta$_3$As and Ta$_3$Bi, while they should be observable in all other compounds investigated in this paper.

\begin{figure}[t!]
\centering
\includegraphics[width=1\linewidth]{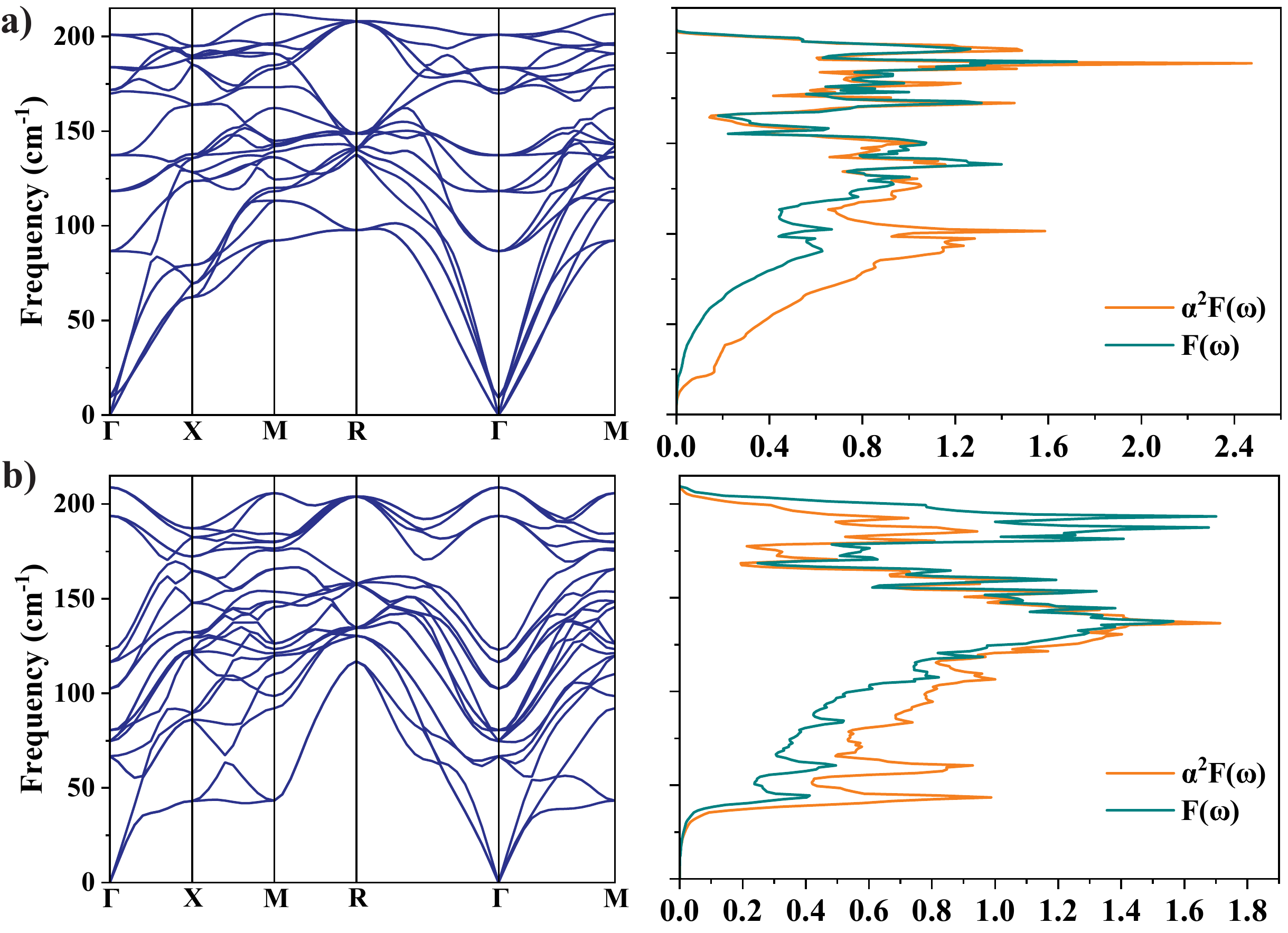}
\caption{Phonon dispersion curves of (a) Nb$_3$Ge and (b) Nb$_3$Sn alongside the corresponding anisotropic Migdal-Eliashberg spectral functions ($\alpha^2$F($\omega$)) and phonon density of states (F($\omega$)).}
\label{phonon-nb3ge-nb3sn}
\end{figure}

\subsection{Topological properties of Nb$_3$X (X = Ge, Sn, Sb): effects of Coulomb repulsion}


Although $\mathbb{Z}_2$ invariants are not well defined for Nb$_3$X compounds in the absence of Coulomb repulsion (U), from the band structures, we can see the Dirac surface states as shown in Fig. \ref{bands-nb3y}(d,g) for Nb$_3$Ge, \ref{bands-nb3y}(e,h) for Nb$_3$Sn and \ref{bands-nb3y}(f,i) for Nb$_3$Sb.
To investigate further, we perform DFT + U calculations for Nb$_3$Ge, Nb$_3$Sn, and Nb$_3$Sb. The band structures within DFT + U are shown in Figs. \ref{bands_Nb3Ge}, \ref{bands_Nb3Sn} and \ref{bands_Nb3Sb}.
The Coulomb repulsion opens a global gap in the momentum space which is obvious from the evolution of bulk band structures for different values of U as shown in Figs. \ref{bands_Nb3Ge}(a) for Nb$_3$Ge, \ref{bands_Nb3Sn}(a) for Nb$_3$Sn and \ref{bands_Nb3Sb}(a) for Nb$_3$Sb. The corresponding surface states are instead shown in Figs. \ref{bands_Nb3Ge}(b) for Nb$_3$Ge, \ref{bands_Nb3Sn}(b) for Nb$_3$Sn and \ref{bands_Nb3Sb}(b) for Nb$_3$Sb. As we can see from the surface states, for all three compounds, the Dirac point at $\Gamma$ is buried in the bulk at U = 0 eV, while already at U=2 eV it is clearly visible and at U = 4 eV there is a global gap in the momentum space making the calculation of four $\mathbb{Z}_2$ invariants possible.

After the opening of the gap, the calculation of $\mathbb{Z}_2$ is well-defined and we obtain the four $\mathbb{Z}_2$ topological invariants ($\nu_0$,$\nu_1\nu_2\nu_3$)=(1;000), showing that all these Nb-based compounds are $\mathbb{Z}_2$ topological metals hosting Dirac surface states similarly to Ta$_3$Sb compound. This is a clear signature of non-trivial topological states appearing not only in heavy Ta-based compounds but also in Nb-based compounds.

\begin{figure*}[t!]
\centering
\includegraphics[width=0.85\linewidth]{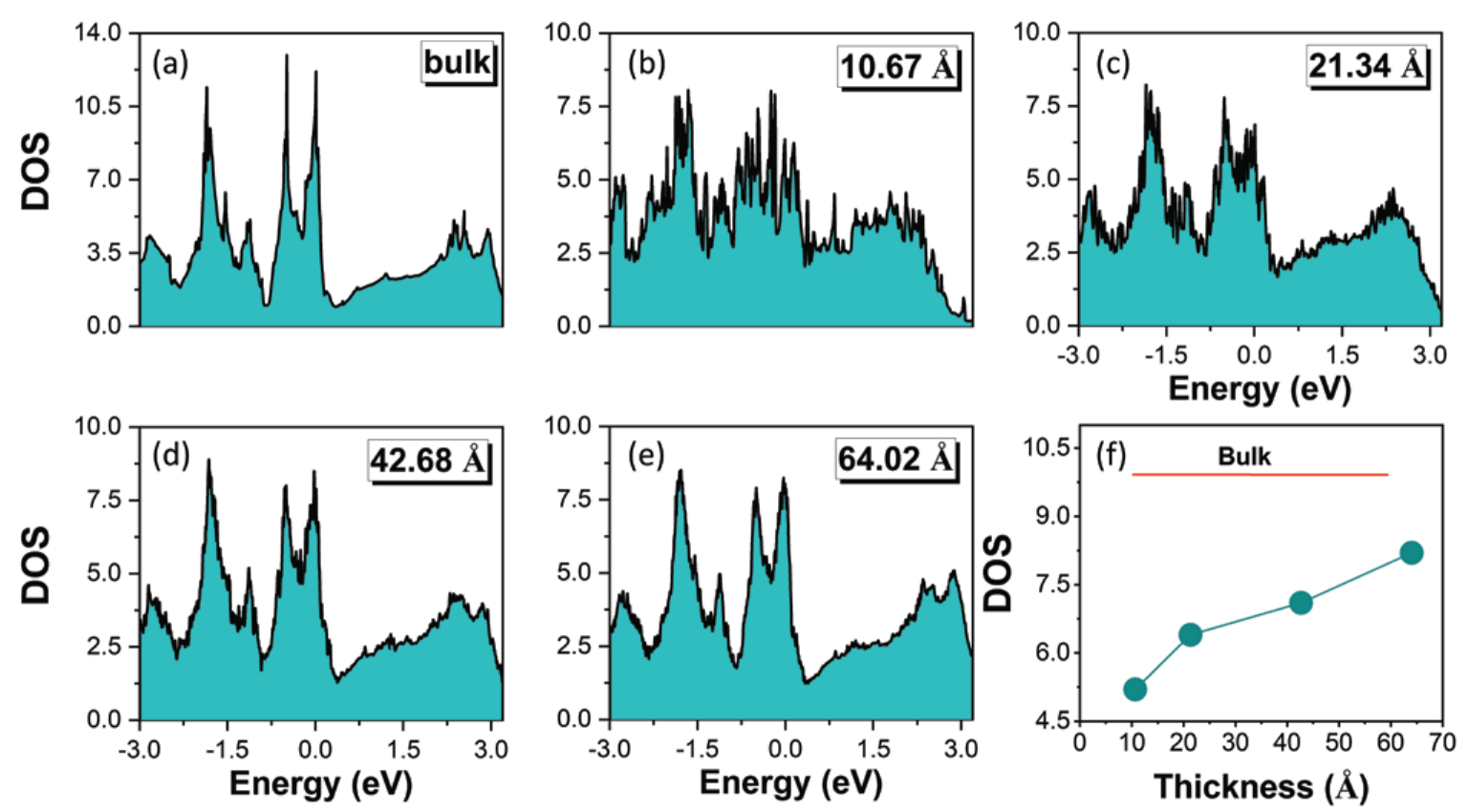}
\caption{Total density of states (DOS) for bulk Nb$_3$Sn (a) and for various thicknesses (b,c,d,e). The Fermi level is set to zero in all panels highlighted in red dashed line. The trend of increasing DOS as a function of thickness is shown in panel (f). The red line represents the DOS value of the bulk. 
}
\label{Thickness_DOS}
\end{figure*}

\subsection{Bulk superconductivity in Nb$_3$X and Ta$_3$Y}

Generally, superconductivity requires metallic states at the Fermi level whereas topological insulators are gaped due to spin-orbit interactions, and they present conducting surface states. Hence, finding a stoichiometric composition where bulk superconductivity and topological surface states coexist is a tough task i.e., the surface states should lie at the Fermi level while the bulk remains fully gaped superconductor in the critical temperature regime. Typically in such systems the Dirac points exist farther from the Fermi level in the conduction bands making it challenging to be observed in experiments like Angle-resolved photoemission spectroscopy  (ARPES).

Several compounds have been investigated to this effect with A15 compounds not being an exception due to their metallic character.\cite{derunova2019giant,Kim19} Studies have been dedicated to Ta$_3$Sb as a potential candidate for topological superconductivity due to the presence of well-resolved spin-orbit induced band manifolds and a superconducting critical temperature of 0.7 K.\cite{Kim19} We revisited this compound and find that in agreement with the previous studies, the Dirac dispersions in surface states are around 500 meV away from the Fermi level in the conduction bands as presented in Fig \ref{bands-ta3x}(e) with a superconducting critical temperature of 0.81 K. However, these Dirac dispersions on the top surface merge with the s-bands of the pnictogens and on the bottom surface merge with the d-bands of Ta at the $\Gamma$ point in the momentum space. Hence it is highly unlikely that Ta$_3$Sb will become a topological superconductor as has been observed in some noncentrosymmetric binary compound BiPd where the Dirac dispersions lie away from the Fermi level in the superconducting regime. This explanation holds true for Ta$_3$As and Ta$_3$Bi as well which have similar characteristics on the surface states (presented in Fig. \ref{bands-ta3x} (d,f)) with superconducting critical temperatures of 3.0 K and 1.16 K respectively.
The s-bands in the conduction band minimum (CBM) of Ta$_3$Bi and Ta$_3$As are responsible for higher $T_c$ as compared to Ta$_3$Sb since they increase the density of states near the Fermi level. 

On the other hand, the $T_c$ for Nb$_3$Ge and Nb$_3$Sn compounds is higher as compared to Ta$_3$Y and Nb$_3$Sb. Owing to the electronic structure near the Fermi level, Nb$_3$Sb has a $T_c$ of 2.21 K which is comparable to that of Ta$_3$As and Ta$_3$Bi. However, Nb$_3$Sb is an outlier when compared to Nb$_3$Ge and Nb$_3$Sn which have a $T_c$ of 15.25 K and 15.66 K respectively. This distinction originates from the lattice dynamics which is evident from the phonon dispersion curves. As compared to Nb$_3$Sn which exhibits anomalous vibrational properties such as soft modes in the $\Gamma \rightarrow$ X $\rightarrow$ M directions in the momentum space (as presented in Fig. \ref{phonon-nb3ge-nb3sn}(b)), the phonon softening is not observed in Nb$_3$Sb presented in Fig. \ref{phonon-ta3y-nb3sb}(d) (see Appendix D). The anomaly of longitudinal acoustic modes of Nb$_3$Sn softening at lower temperature scales is accompanied by large neutron scattering linewidths which is a function of the electron-phonon coupling coefficient $\lambda$($\omega$), hence resulting in higher $T_c$.\cite{Stewart15}

We present the phonon dispersion curves alongside the Migdal-Eliashberg spectral functions $\alpha^2$F($\omega$) and phonon density of states F($\omega$) for Nb$_3$Ge and Nb$_3$Sn are presented in Fig. \ref{phonon-nb3ge-nb3sn} and for Ta$_3$Y, Nb$_3$Sb are presented in Fig. \ref{phonon-ta3y-nb3sb} (see Appendix D). The electron-phonon coupling coefficients $\lambda$($\omega$) for Nb$_3$Ge and Nb$_3$Sn are 1.41 and 2.03, respectively. The bulk superconductivity in Nb$_3$X compounds is quite sensitive to the method of crystal growth and various experimental conditions vary the experimental $T_c$. However, Nb$_3$Ge has been found to exhibit a maximum $T_c$ of 23.2 K which along with Nb$_3$Sn gives further scope to explore the interplay between the topology and superconductivity owing to large SHC at the Fermi level.

\subsection{Thickness dependence of the electronic and superconducting properties in Nb$_3$Sn thin films}

We calculate the DOS as a function of the thickness to asses the superconducting properties of the Nb$_3$Sn thin film.
The electronic properties reported in this subsection were calculated with the computational framework described in Appendix C. In Fig. \ref{Thickness_DOS}(a-e), we show the DOS calculated for the bulk and for the stoichiometric slabs with different thicknesses. The DOS as a function of the thickness is reported in Fig. \ref{Thickness_DOS}(f), we observe a decreasing value of the DOS when the thickness got reduced, this will reflect in a reduction of the critical temperature for the thin films of Nb$_3$Sn. For ultra-thin films, the DOS is reduced approximately by half with respect to the bulk value. Consequently, the observed decline in superconductivity in films with reduced thickness can be explained by a weakening of the DOS. While the large DOS in three dimensions can be attributed to the presence of the van Hove singularities\cite{Labb1967}, in two dimensions we observed a smoothing of the van Hove singularity.  
These calculations show that Nb$_3$Sn has a different behavior due to its complex band structure properties compared to other BCS superconductors like V, Ta or Nb where one normally finds that, as soon as the thin film thickness is equal or greater than the coherence length, $T_c$ is very close to the bulk $T_c$. 
The experimental superconducting critical temperature of the thin films strongly depends on the sample quality, of the substrate and usually tends to reduce with the thickness reduction.\cite{ILIN2010953,Asano88}

\section{Tight-binding model with three SSH chains for \lowercase{t$_{2g}$} orbitals}

Here, we report a model that includes only the t$_{2g}$ orbitals of the Nb/Ta atoms. This tight-binding model allows us to understand the character of the orbitals and the Fermi level and which hopping parameters tune the opening of the topological gap.
If we want to produce a minimal model for the d-orbitals of the Nb/Ta atoms, we must include all the 6 Nb/Ta atoms per unit cell.
If we could decouple p- and d-electrons, e$_g$ are below in energy with respect to t$_{2g}$ given the crystal field due to the position of other atoms, or anyway, we can assume this in first approximation. Since the charge transfer is zero, the Nb/Ta atoms are in d$^5$ electronic configuration. We developed a tight-binding model for the t$_{2g}$ subset for the 6 Nb/Ta-atoms. Within the t$_{2g}$ tight-binding model we can easily include the spin-orbit coupling. 
The crystal structure presents three dimers of Nb/Ta atoms, along the $a$, $b$ and $c$ axes, as shown in Fig. \ref{Crystal structure}a). We consider in our model the intradimer hybridizations and the interdimer hoppings by including the first and the second nearest neighbours.

\begin{figure}[t!]
\centering
\includegraphics[width=\linewidth]{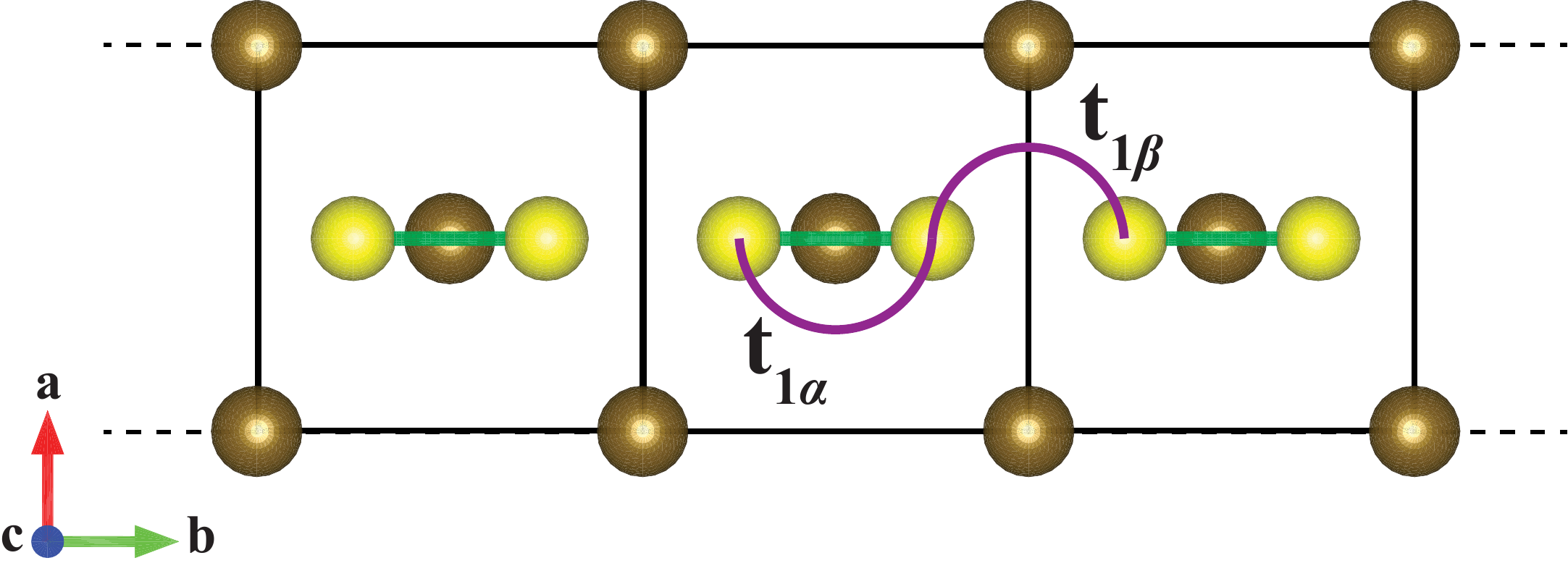}
\caption{Top view of the crystal structure of the A$_3$B compound. The $A$ sites are shown in yellow and $B$ sites are shown in brown. We indicate with a green bond the dimer along the b-axis, we have deleted the other dimers for a better visualization. The hopping parameters t$_{1\alpha}$ and t$_{1\beta}$ have the same geometric structure as the hopping in the SSH chain.}
\label{SSH}
\end{figure}

\begin{figure}[t!]
\centering
\includegraphics[width=0.85\linewidth]{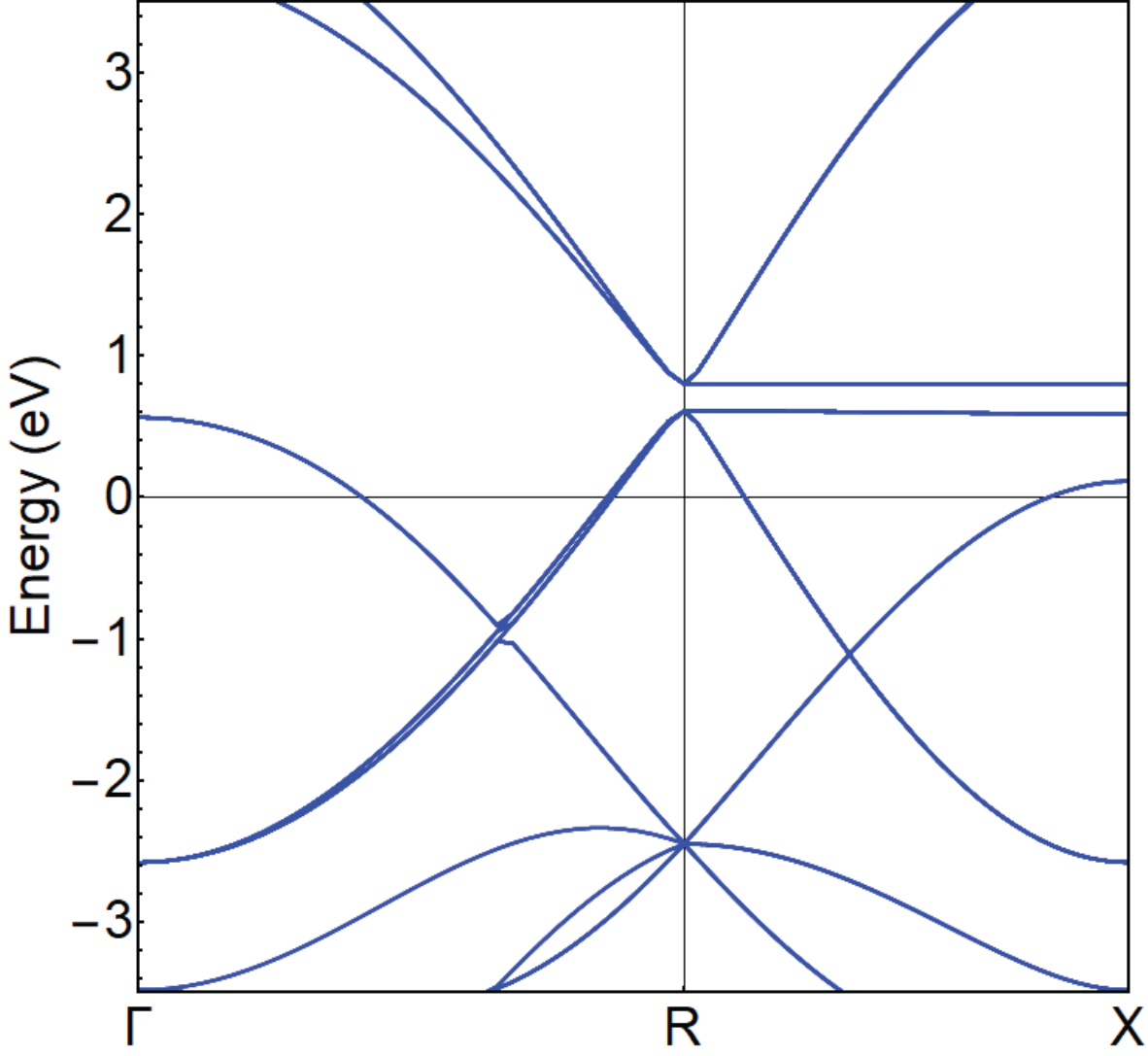}
\caption{Band structure reproduced with the minimal model without including the SOC interaction. The parameters used are: E$_1$=6.359 eV, E$_2$=3.205 eV, t$_{1\alpha}$=1.691 eV, t$_{1\beta}$=1.594 eV, t$_{2\alpha}$=$-$ 0.516 eV, t$_{2\beta}$=$-$ 0.518 eV, t$_{3}$=$-$0.065 eV and t$_{4}$=$-$0.495 eV. The value of the onsite energy E$_1$ controls the position of the bands that are at around 1 eV at the R point, while the value of E$_2$ controls the position of the bands manifold which is at around -3 eV at the R point. The Fermi level is set at zero energy. 
}
\label{Bands_NO_SOC_TB}
\end{figure}

\begin{figure}[t!]
\centering
\includegraphics[width=\linewidth]{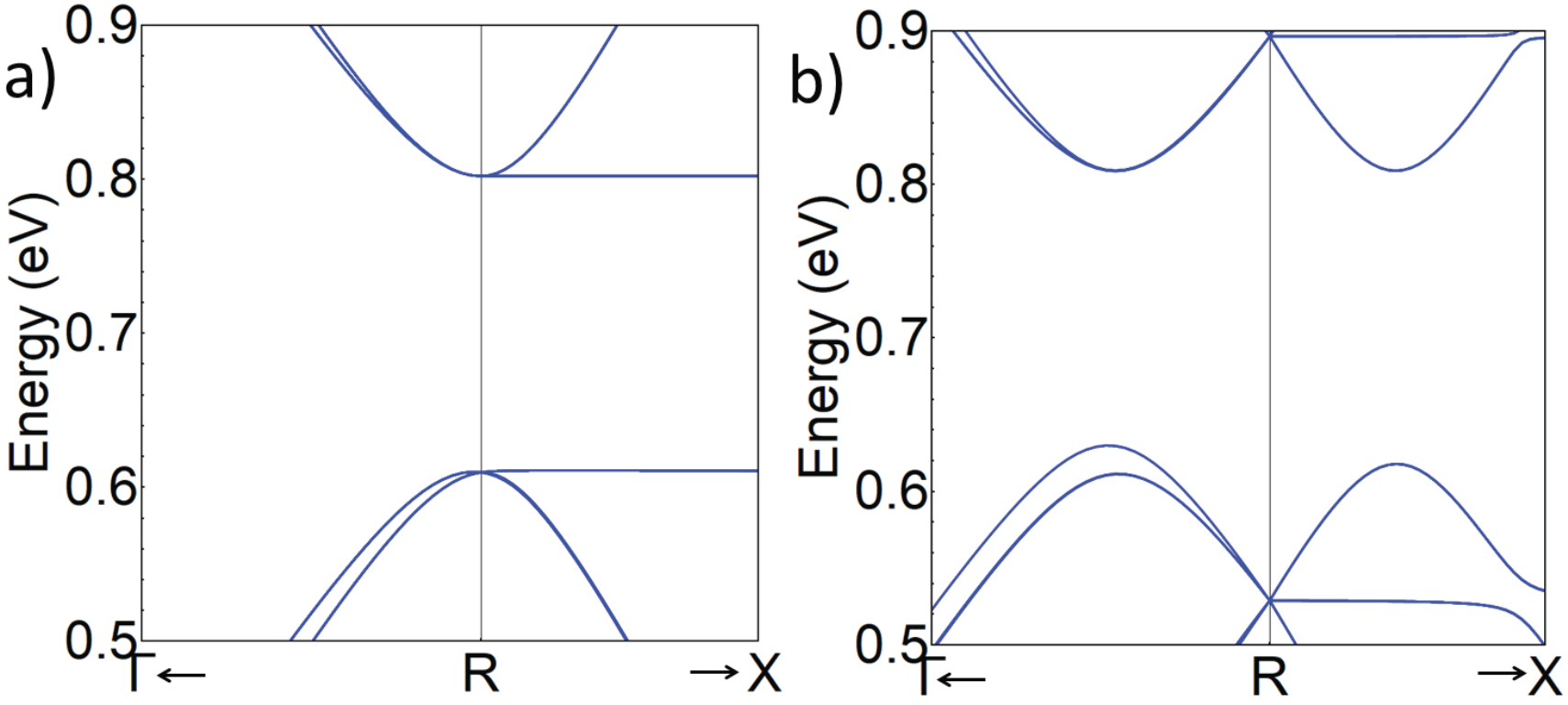}
\caption{Zoom of the band structure reproduced with the minimal model along the lines $\Gamma$-R-X a) without SOC and b) with the SOC interaction included. The parameters used are: E$_1$=6.359 eV, E$_2$=3.205 eV, t$_{1\alpha}$=1.691 eV, t$_{1\beta}$=1.594 eV, t$_{2\alpha}$=$-$ 0.516 eV, t$_{2\beta}$=$-$ 0.518 eV, t$_{3}$=$-$0.065 eV and t$_{4}$=$-$0.495 eV. The SOC parameter is $\lambda$=0.3 eV. The energy difference between the bands at the R point without SOC is roughly proportional to the module of the difference between the hoppings t$_{1\alpha}$ and t$_{1\beta}$. The Fermi level is set at zero energy. 
}
\label{Bands_SOC_zoom_TB}
\end{figure}

The spinful tight-binding model is reported in more detail in Appendix A and it is composed of three coupled SSH chains with t$_{2g}$ orbitals. The SSH chain along the b-axis is shown in Fig. \ref{SSH}. Very few recent examples of a three-dimensional SSH model have been reported in the literature\cite{PhysRevB.107.045135,PhysRevB.107.125123}, with significant differences from the case proposed in this paper.
The parameters of the model are described in detail in Appendix A, we have two onsite energies E$_1$ and E$_2$ for the $\pi$ and $\delta$-bonds, respectively. The hopping parameters t$_{1\alpha}$, t$_{1\beta}$, t$_{2\alpha}$, t$_{2\beta}$ for the intradimer elements and t$_{3}$ and t$_{4}$ for the interdimer elements. 
The band structure reproduced of the model without including SOC is reported in Figs. \ref{Bands_NO_SOC_TB}.
A magnification of the band structure at the R point is shown in \ref{Bands_SOC_zoom_TB}a) and \ref{Bands_SOC_zoom_TB}b), without and with SOC respectively.
We include the SOC within the t$_{2g}$ as the SOC of L=$-$1.
In our model, when t$_{1\alpha}$=t$_{1\beta}$ and t$_{2\alpha}$=t$_{2\beta}$, the $\pi$ bands are 24 times degenerate without SOC, while when SOC is applied, they are splitted in two sets 12 times degenerate. The $\delta$ bands are 12 times degenerate both without and with SOC. In the realistic case, namely when t$_{1\alpha}$ is different from t$_{1\beta}$ and t$_{2\alpha}$ is different from t$_{2\beta}$, the $\pi$ bands are 12 times degenerate without SOC while with SOC they are six-fold degerate. The $\delta$ bands are six-fold degenerate both without and with SOC. When t$_{2\alpha}$=t$_{2\beta}$ and t$_{1\alpha}$ is different from t$_{1\beta}$, the $\pi$ bands are 12 times degenerate and the $\delta$ bands are 12 times degenerate both without SOC and with SOC. When t$_{1\alpha}$=t$_{1\beta}$ and t$_{2\alpha}$ is different from t$_{2\beta}$, the $\pi$ are 24 times degenerate without SOC and 12 times degenerate with SOC, while the the $\delta$ bands are 12 times degenerate without SOC and six times degenerate with SOC.
The opening of the topological gap between the conduction and the valence band at R is controlled by both the difference t$_{1\alpha}$-t$_{1\beta}$ and the spin-orbit coupling, also the Coulomb repulsion controls the gap as it was proved within DFT+U. The opening of the gap at R allows us to explicitly calculate the topological invariants.
The SSH model is a spinless and chiral 1D model, while the model that we have proposed is spinful and not chiral. Despite these differences regarding the material class, the relevant quantity for the topological properties is the difference between the hopping parameters due to the dimerization in both cases.

\section{Discussion, conclusions and outlook}

Ta$_3$Sb shows robust orbital textures in the topological surface states\cite{Kim19}. Since the material class is the same and topological surface states are qualitatively the same, we expect the orbital texture even in Nb$_3$Ge and Nb$_3$Sn. The proximity-induced superconductivity in the Dirac surface states can generate Majorana zero mode localized at the vortex\cite{Kim19,PhysRevLett.100.096407}. Further studies using the tight-binding model for the bulk and the surface of A15 could shed light on the interplay between topological and superconductive properties.\\

In conclusion, using a first-principle approach, we provide an extensive description of the electronic, topological and superconductive properties of the Nb- and Ta-based A15 compounds by calculating band structure, SHC, superconducting $T_c$, and topological surface states. 
All compounds Nb$_3$X (X = Ge, Sn, Sb) and Ta$_3$Y (Y = As, Sb, Bi) have metallic band structures. Nb$_3$Ge and Nb$_3$Sn have one electron less respect to the other compounds, they have a larger density of states and high superconducting $T_c$. Ta$_3$As and Ta$_3$Bi host the Ta-6s band at the Fermi level producing larger $T_c$ and larger SHC compared to Ta$_3$Sb due to the presence of additional DOS and additional anticrossings close to the Fermi level.
The spin-Hall conductivity is relatively large also for the lighter elements due to the presence of several anticrossings in the Brillouin zone. All the superconducting $T_c$ are sizeable.
One of the most interesting results among our outcomes is the presence of Dirac surface states at the $\Gamma$ point for all compounds at the same filling. The ideal case is the Ta$_3$Sb where we have $\mathbb{Z}_2$ metallic compounds with net separation between conduction and valence bands.
In all the other compounds, the conduction and the valence bands cross due to the Ta-6s for the Ta-compounds and due to the weaker SOC in the Nb-compounds. Despite the crossing between conduction and valence bands, the band inversion at the high-symmetry points persists in Nb-based compounds producing always the Dirac surface states at the $\Gamma$ point. Unfortunately, the Dirac surface states are obstructed when we include the hybridization with the s-bands in the Ta-based compounds. 
The surface Dirac points can be tuned by the Coulomb repulsion, in the case of transition metal termination the Dirac point is around the Fermi level. 
Even if the $\mathbb{Z}_2$ topological invariant cannot be calculated in all cases, the presence of the Dirac surface states is persistent. Therefore, we can assume that the Nb$_3$X (X = Ge, Sn, Sb) and Ta$_3$Y (Y = As, Sb, Bi) compounds are all $\mathbb{Z}_2$ topological metals.
These Dirac surface states could explain the low resistivity of the Nb$_3$Sn surface observed experimentally\cite{10.1063/5.0015376}.
Additionally, we provide a minimal tight-binding model composed of three coupled SSH chains and based on the t$_{2g}$ orbitals. This tight-binding reproduces the relevant electronic and topological features for these compounds at the Fermi level. With a $T_c$ of 23.2 K, Nb$_3$Ge could be the $\mathbb{Z}_2$ topological metal with the highest $T_c$ ever reported amongst the A15 compounds.\\

The surfaces of the A15 will host an interplay between $\mathbb{Z}_2$ topology, robust orbital texture, breaking of the inversion symmetry and BCS superconductivity with relatively large $T_c$. The non-trivial topology originates from the well resolved bands in the momentum space which dues to bulk-boundary correspondence host Dirac dispersions along the surfaces. Also, alongside this topological phenomena, due to multiple bands crossing the Fermi energy, the superconductivity is retained simultaneously. Therefore, the surfaces of A15 are a platform to search for exotic topological superconductivity. Once it will be grown the thin film of A15, it will be possible to construct superlattices, junctions, or heterostructures of superconductors and topological compounds in order to study the topological superconductivity via the proximity effect. The non-trivial surface properties of A15 thin films can also represent an interesting platform for the realization of gate-controllable superconducting devices, where recent studies have suggested that surface properties are key for the observation of the suppression of a critical current under an applied gate voltage \cite{Ruf23}.

\section*{Acknowledgments}
We acknowledge B. Wieder for useful discussions.
The work is supported by the Foundation for Polish Science through the International Research Agendas program co-financed by the European Union within the Smart Growth Operational Programme (Grant No. MAB/2017/1). A.D.B. and M.C. acknowledge partial support by the EU’s Horizon 2020 Research and Innovation Framework Program under Grant Agreement No. 964398 (SUPERGATE).
C. A. acknowledges Erasmus+ for a training scholarship.
We acknowledge the access to the computing facilities of the Interdisciplinary Center of Modeling at the University of Warsaw, Grant g91-1418, g91-1419 and g91-1426 for the availability of high-performance computing resources and support. We acknowledge the CINECA award under the ISCRA initiative  IsC99 "SILENTS”, IsC105 "SILENTSG", IsB26 "SHINY" and IsB27 "SLAM" grants for the availability of high-performance computing resources and support. We acknowledge the access to the computing facilities of the Poznan Supercomputing and Networking Center Grant No. 609 (pl0223-01) and pl0267-01.

\appendix

\section{Three coupled SSH chains with a t$_{2g}$ orbital basis}

Our model includes just the t$_{2g}$ electrons for the 6 Nb-atoms. The crystal structure presents three dimers of Nb atoms, along the $a$, $b$ and $c$ axes. We consider in our model the intradimer hybridizations and the interdimer hoppings by considering the first and the second nearest neighbours.

The lattice constant is d=5.139 {\AA}.
The basis in the Hilbert space is given by the vector 
\begin{equation}
\phi_{i}^{\dagger}=(\phi_{1a}^{\dagger},\phi_{2a}^{\dagger},\phi_{3c}^{\dagger},\phi_{4c}^{\dagger},\phi_{5b}^{\dagger},\phi_{6b}^{\dagger}),\label{eqn:vectors1}
\end{equation}

where 
\begin{equation}
\phi_{1a}^{\dagger}=(\phi_{1a,xz}^{\dagger},\phi_{1a,yz}^{\dagger},\phi_{1a,xy}^{\dagger}),\label{eqn:vectors2}
\end{equation}

the indices 1,.., 6 indicate the Nb atoms from the first to the sixth, the letters $a$, $c$, $b$ indicate the fact that these sites belong to the dimers along the cell directions $a$, $c$ and $b$, while xz, yz and xy indicate the t$_{2g}$ orbitals d$_{xz}$, d$_{yz}$ and d$_{xy}$.

Our Hamiltonian is the following:

\[
H=\begin{bmatrix}H_{aa} & H_{ac} & H_{ab}\\
H_{ca} & H_{cc} & H_{cb}\\
H_{ba} & H_{bc} & H_{bb}
\end{bmatrix},
\]

where the subscript $a$ indicates the dimer along the $a$ direction, composed of Nb$_1$ and Nb$_2$ atoms,  $c$ the dimer along the $c$ direction, composed of Nb$_3$ and Nb$_4$ atoms and $b$ indicates the dimer along the $b$ direction, namely the dimer of Nb$_5$ and Nb$_6$ atoms.
The $aa$, $bb$ and $cc$ blocks of the Hamiltonian contain the intradimer hybridizations, while the off-diagonal blocks contain the interdimer hoppings.

The intradimer submatrices are of this type:

\[
H_{aa}=\begin{bmatrix}H_{1a1a} & H_{1a2a} \\
H_{2a1a} & H_{2a2a}
\end{bmatrix},
\]

where

\[
H_{1a1a}=\begin{bmatrix}E_{1a_{xz}} & 0 & 0 \\
0 & E_{1a_{yz}} & 0 \\
0 & 0 & E_{1a_{xy}} \\
\end{bmatrix},
\]

and 

\[
H_{1a2a}=\begin{bmatrix}H_{1a_{xz}2a_{xz}} & 0 & 0 \\
0 & H_{1a_{yz}2a_{yz}} & 0 \\
0 & 0 & H_{1a_{xy}2a_{xy}} \\
\end{bmatrix},
\]

The on-site energies belong to two groups. The energy $E_{1}$ belongs to the orbitals that form intradimer $\pi$-bond, while the energy $E_{2}$ belongs to the orbitals that form intradimer $\delta$-bond :

\begin{equation*}
\begin{split}
E_{1a_{xz}} =  E_{1} \qquad  E_{1a_{yz}}  =  E_{2} \qquad  E_{1a_{xy}}  =  E_{1} \\
E_{2a_{xz}} =  E_{1} \qquad  E_{2a_{yz}}  =  E_{2} \qquad  E_{2a_{xy}}  =  E_{1} \\
E_{3c_{xz}} =  E_{1} \qquad  E_{3c_{yz}}  =  E_{1} \qquad  E_{3c_{xy}}  =  E_{2} \\
E_{4c_{xz}} =  E_{1} \qquad  E_{4c_{yz}}  =  E_{1} \qquad  E_{4c_{xy}}  =  E_{2} \\
E_{5b_{xz}} =  E_{2} \qquad  E_{5b_{yz}}  =  E_{1} \qquad  E_{5b_{xy}}  =  E_{1} \\
E_{6b_{xz}} =  E_{2} \qquad  E_{6b_{yz}}  =  E_{1} \qquad  E_{6b_{xy}}  =  E_{1}
\end{split}
\end{equation*}
The intradimer elements have a hopping form similar to the SSH model\cite{PhysRevLett.42.1698} as we can see in Fig. \ref{SSH}, therefore, the tight-binding model that describes the A15 is composed of three coupled SSH chains.
The intradimer Hamiltonian elements have the following form:
\begin{eqnarray*}
H_{1a_{xz}2a_{xz}} & = &  t_{1\alpha}e^{-ik_xd/2}  + t_{1\beta}e^{ik_xd/2} \\ 
H_{1a_{yz}2a_{yz}} & = &  t_{2\alpha}e^{-ik_xd/2}  + t_{2\beta}e^{ik_xd/2} \\
H_{1a_{xy}2a_{xy}} & = &  t_{1\beta}e^{-ik_xd/2}  + t_{1\alpha}e^{ik_xd/2} \\
H_{3c_{xz}4c_{xz}} & = &  t_{1\beta}e^{-ik_xd/2}  + t_{1\alpha}e^{ik_xd/2} \\ 
H_{3c_{yz}4c_{yz}} & = &  t_{1\alpha}e^{-ik_xd/2}  + t_{1\beta}e^{ik_xd/2} \\
H_{3c_{xy}4c_{xy}} & = &  t_{2\alpha}e^{-ik_xd/2}  + t_{2\beta}e^{ik_xd/2} \\
H_{5b_{xz}6b_{xz}} & = &  t_{2\alpha}e^{-ik_xd/2}  + t_{2\beta}e^{ik_xd/2} \\ 
H_{5b_{yz}6b_{yz}} & = &  t_{1\alpha}e^{-ik_xd/2}  + t_{1\beta}e^{ik_xd/2} \\
H_{5b_{xy}6b_{xy}} & = &  t_{1\beta}e^{-ik_xd/2}  + t_{1\alpha}e^{ik_xd/2}
\end{eqnarray*}
where t$_1$ are $\pi$-bonds hoppings and t$_2$ are $\delta$-bonds hoppings. As expected, we have t$_1$ $>$ t$_2$ if we extract the parameters from the wannierization of the DFT band structure. The configuration of $\alpha$ and $\beta$ are the left and right hopping, their configuration is related to the symmetries of the system. The topological gap at the R-point is controlled by the difference between t$_{1\alpha}$ - t$_{1\beta}$ and enhanced by the spin-orbit coupling.

The interdimer submatrices are of the type:
\[
H_{ac}=\begin{bmatrix}H_{1a3c} & H_{1a4c} \\
H_{2a3c} & H_{2a4c}
\end{bmatrix},
\]
where we have:
\[
H_{1a3c}=\begin{bmatrix}H_{1a_{xz}3c_{xz}} & H_{1a_{xz}3c_{yz}} & H_{1a_{xz}3c_{xy}} \\
H_{1a_{yz}3c_{xz}} & H_{1a_{yz}3c_{yz}} & H_{1a_{yz}3c_{xy}} \\
H_{1a_{xy}3c_{xz}} & H_{1a_{xy}3c_{yz}} & H_{1a_{xy}3c_{xy}} \\
\end{bmatrix},
\]
Regarding the interdimer hybridizations, the intraorbital elements that are different from zero in our model are the following:
\begin{eqnarray*}
H_{1a_{xz}3c_{xz}} & = &  2t_{3}e^{i(k_x+k_z)d/4}cos(k_yd/2) \\ 
H_{1a_{xz}4c_{xz}} & = &  2t_{3}e^{i(k_x-k_z)d/4}cos(k_yd/2) \\ 
H_{1a_{xy}5b_{xy}} & = &  2t_{3}e^{i(-k_x+k_y)d/4}cos(k_zd/2) \\ 
H_{1a_{xy}6b_{xy}} & = &  2t_{3}e^{i(-k_x-k_y)d/4}cos(k_zd/2) \\ 
H_{2a_{xz}3c_{xz}} & = &  2t_{3}e^{i(-k_x+k_z)d/4}cos(k_yd/2) \\ 
H_{2a_{xz}4c_{xz}} & = &  2t_{3}e^{i(-k_x-k_z)d/4}cos(k_yd/2) \\ 
H_{2a_{xy}5b_{xy}} & = &  2t_{3}e^{i(k_x+k_y)d/4}cos(k_zd/2) \\ 
H_{2a_{xy}6b_{xy}} & = &  2t_{3}e^{i(k_x-k_y)d/4}cos(k_zd/2) \\
H_{3c_{yz}5b_{yz}} & = &  2t_{3}e^{i(-k_y+k_z)d/4}cos(k_xd/2) \\
H_{3c_{yz}6b_{yz}} & = &  2t_{3}e^{i(k_y+k_z)d/4}cos(k_xd/2) \\
H_{4c_{yz}5b_{yz}} & = &  2t_{3}e^{i(-k_y-k_z)d/4}cos(k_xd/2) \\
H_{4c_{yz}6b_{yz}} & = &  2t_{3}e^{i(k_y-k_z)d/4}cos(k_xd/2) \\
\end{eqnarray*}
and the interorbital Hamiltonian elements are the following: 
\begin{eqnarray*}
H_{1a_{yz}3c_{xy}} & = &  2t_{4}e^{i(k_x+k_z)d/4}cos(k_yd/2) \\
H_{1a_{yz}4c_{xy}} & = &  -2t_{4}e^{i(k_x-k_z)d/4}cos(k_yd/2) \\
H_{1a_{yz}5b_{xz}} & = &  -2t_{4}e^{i(-k_x+k_y)d/4}cos(k_zd/2) \\ 
H_{1a_{yz}6b_{xz}} & = &  2t_{4}e^{i(-k_x-k_y)d/4}cos(k_zd/2) \\ 
H_{2a_{yz}3c_{xy}} & = &  -2t_{4}e^{i(-k_x+k_z)d/4}cos(k_yd/2) \\ 
H_{2a_{yz}4c_{xy}} & = &  2t_{4}e^{i(-k_x-k_z)d/4}cos(k_yd/2) \\ 
H_{2a_{yz}5b_{xz}} & = &  2t_{4}e^{i(k_x+k_y)d/4}cos(k_zd/2) \\ 
H_{2a_{yz}6b_{xz}} & = &  -2t_{4}e^{i(k_x-k_y)d/4}cos(k_zd/2) \\
H_{3c_{xy}5b_{xz}} & = &  -2t_{4}e^{i(-k_y+k_z)d/4}cos(k_xd/2) \\
H_{3c_{xy}6b_{xz}} & = &  2t_{4}e^{i(k_y+k_z)d/4}cos(k_xd/2) \\
H_{4c_{xy}5b_{xz}} & = &  2t_{4}e^{i(-k_y-k_z)d/4}cos(k_xd/2) \\
H_{4c_{xy}6b_{xz}} & = &  -2t_{4}e^{i(k_y-k_z)d/4}cos(k_xd/2) \\
\end{eqnarray*}

\begin{figure}[t!]
\centering
\includegraphics[width=0.99\linewidth]{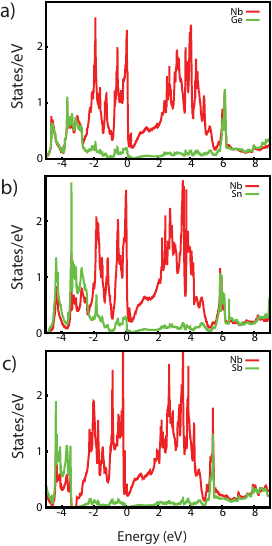}
\caption{Local density of states of (a) Nb$_3$Ge, (b) Nb$_3$Sn and (c) Nb$_3$Sb. The Fermi level is set to zero in all panels.}
\label{DOSGe}
\end{figure}

\section{Computational details for the Quantum Espresso calculations}

To compute material properties, we employed density functional theory based \textit{first-principles} calculations as implemented within the Quantum ESPRESSO code.\cite{giannozzi2009quantum} We use generalized-gradient approximations with Perdew-Burke-Ernzerhof (PBE) exchange-correlation functional \cite{perdew1996generalized} implemented in norm-conserving pseudopotential with optimized kinetic energy cut-off of 110 Ry for Nb$_3$X (X = Ge, Sb, Sn) and 70 Ry Ta$_3$Y (Y = Bi, As, Sb). A uniform Monkhorst-Pack grid of 20 $\times$ 20 $\times$ 20 was used in all the calculations which corresponds to 1342 special \textbf{\textit{k}}-points in the irreducible Brillouin zone. In phonons of Nb$_3$Sn, we observed a numerical artifact at high symmetry point $\Gamma$ with PBE pseudopotential, hence made additional calculations using local-density approximation with Perdew-Wang 91 gradient-corrected functional only for this system.\cite{perdew1992accurate} The optimized kinetic energy cut-off of 70 Ry was used for this purpose.

The dynamical calculations pertaining to vibrational properties have been performed within the density functional perturbation theory \cite{baroni2001phonons} wherein the dynamical matrices were sampled within the irreducible Brillouin zone using a \textbf{\textit{q}}-mesh of at least 3 $\times$ 3 $\times$ 3. The Migdal-Eliashberg spectral functions were calculated using Eq. \ref{1} presented below.\cite{baroni2001phonons}

\begin{equation}\label{1}
    \alpha^2 F(\omega) = \frac{1}{2\pi N(E_F)} \sum_{qv} \frac{\gamma_{qv}}{\omega_{qv}} \delta(\omega - \omega_{qv})
\end{equation}

where, $\gamma_{qv}$ is phonon linewidth and $\omega_{qv}$ is the phonon eigenfrequency. Integrating these spectral functions over the frequencies as presented in Eq. \ref{2}, we obtain the electron-phonon coupling coefficient $\lambda$($\omega$).

\begin{equation}\label{2}
    \lambda(\omega) = 2 \int_{0}^{\infty} \frac{\alpha^2 F(\omega)}{\omega} d\omega
\end{equation}

Following this, the superconducting critical temperature ($T_c$) was calculated using Allen-Dynes modification of the McMillan formula \cite{allen1975transition} presented in Eq. \ref{3} where $\mu^*$ is effective Coulomb repulsion parameter $\omega_{ln}$ is weighted logarithmic average phonon frequencies.

\begin{equation}\label{3}
T_{c} = \frac{\omega_{log}}{1.2} exp\left(-\frac{1.04(1+\lambda)}{\lambda-\mu^*(1+0.62\lambda)}\right)
\end{equation}

To compute the topological surface states, the $\mathbb{Z}_2$ topological invariants and spin Hall conductivity, we used the Wannier90 \cite{mostofi2014updated} and WannierTools codes \cite{wu2018wanniertools}. 
The basis of the tight-binding model was composed of the Wannier functions \cite{marzari1997maximally} obtained from the d-orbitals of Nb or Ta and the p-orbitals of the other atoms.
The atomic orbital configuration is s$^2$d$^3$ for Nb and Ta, but it becomes d$^5$ in crystals, therefore Ta and Nb have half-filled d-orbitals.
The tight-binding model is composed of 60 d-bands and 12 p-bands for a total of 72 bands. The topological gap is observed at half-filling. Of these 36 occupied bands, 6 are mainly p-bands and 30 are mainly d-bands.
The momentum mesh used for calculating spin Hall conductivity ${\sigma }_{\mathit{\text{xy}}}^{\text{spin}z}(\omega)$ was 200 $\times$ 200 $\times$ 200. This was done using the Kubo-Greenwood formula as implemented in Wannier90 \cite{guo2008intrinsic,qiao2018calculation}.

\begin{figure}[ht!]
\centering
\includegraphics[width=\linewidth]{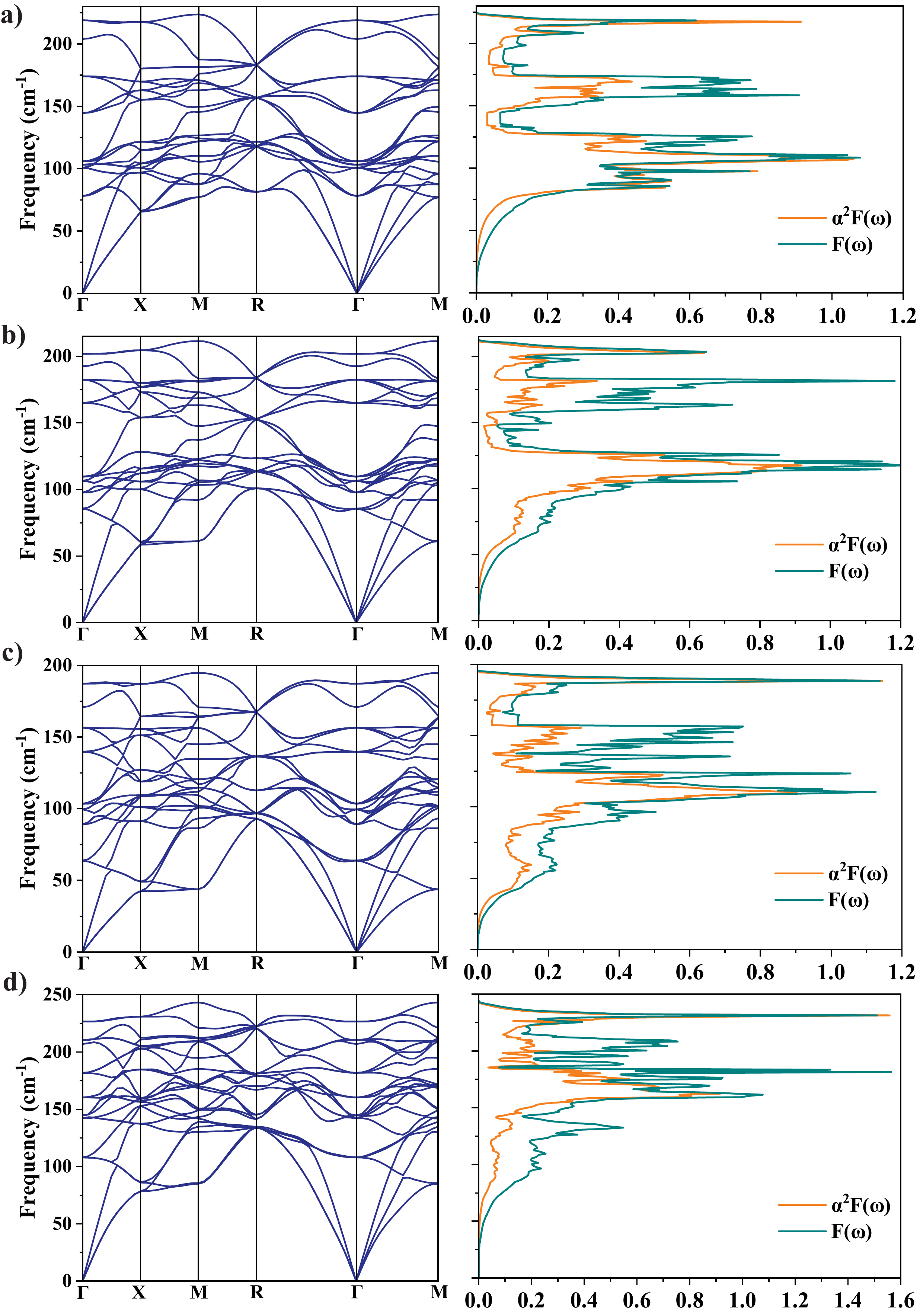}
\caption{Phonon dispersion curves of (a) Ta$_3$As, (b) Ta$_3$Sb, (c) Ta$_3$Bi and (d) Nb$_3$Sb alongside the corresponding anisotropic Migdal-Eliashberg spectral functions $\alpha^2$F($\omega$) and phonon density of states F($\omega$).}
\label{phonon-ta3y-nb3sb}
\end{figure}

\begin{figure*}[ht!]
\centering
\includegraphics[width=\linewidth]{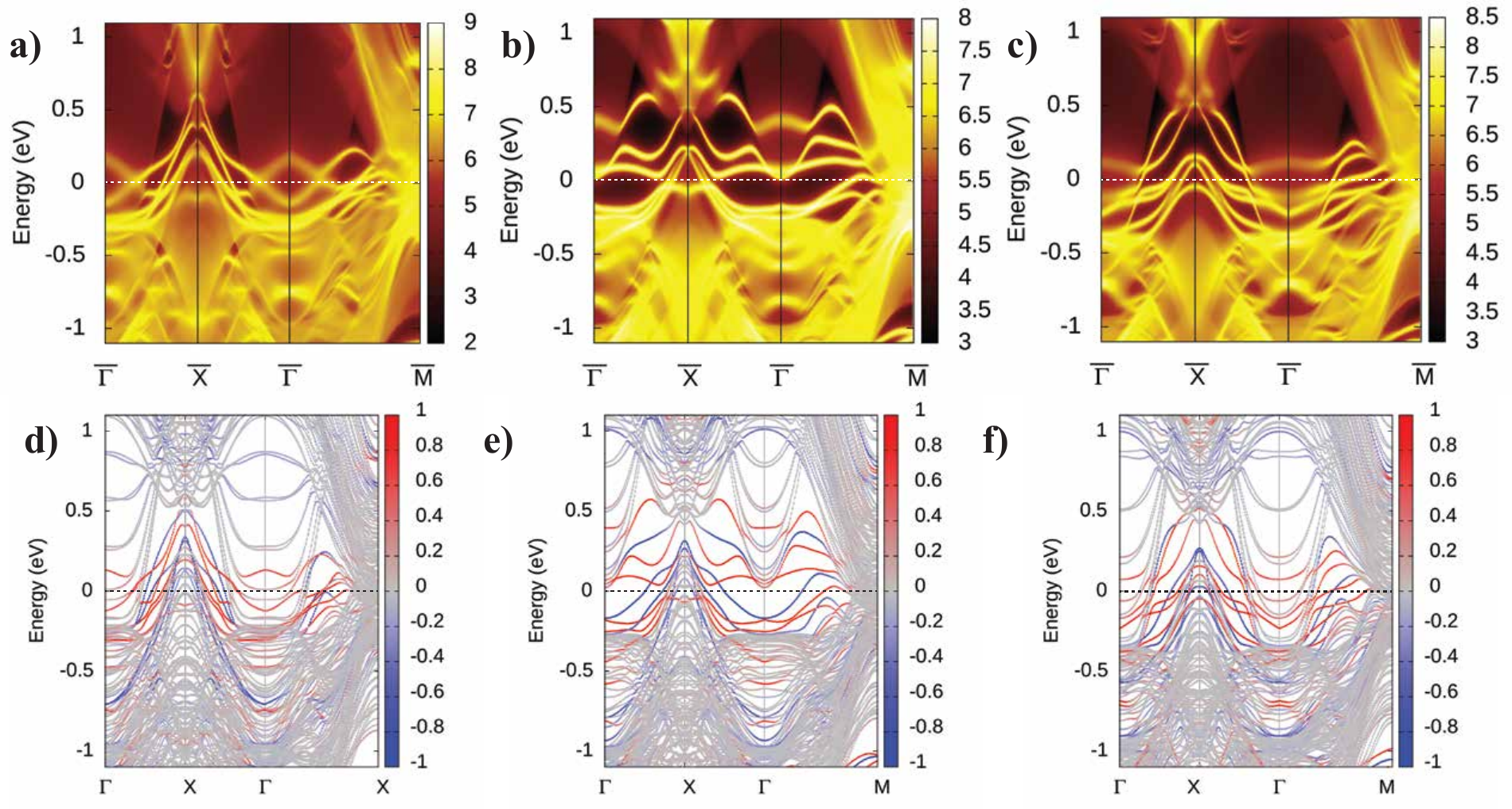}
\caption{Surface states of (a) Ta$_3$As, (b) Ta$_3$Sb and (c) Ta$_3$Bi indicating spin-momentum locked Dirac dispersions at $\Gamma$ point submerged below the s-bands. Slab band structures of (d) Ta$_3$As, (e) Ta$_3$Sb and (f) Ta$_3$Bi with the red bands indicating the contribution from the top surface layer and the blue bands indicating the contribution from the bottom surface layer of the slab presented in Fig \ref{Crystal structure}. The legends on the right in all the panels indicate the spectral weights. The Fermi level is set to zero in all panels.}
\label{with-s-bands-ta3y}
\end{figure*}

\section{Slab calculation and density of states of the Nb-based compounds using VASP}

For the slab calculations, additional first-principles calculations are performed via the Vienna {\it ab initio} simulation package (VASP) \cite{blochl1994projector, kresse1996efficient} using density functional theory. The generalized gradient approximation with PBE form \cite{perdew1996generalized} and PBEsol \cite{perdew2008restoring} is adopted to calculate the lattice parameters and the density of states (DOS). An energy cutoff of 350 eV and a mesh of 16 × 16 × 1 k-points were chosen for the different thicknesses of Nb-based compounds. The calculations were converged with the convergence criteria of 0.01 eV/\AA, and energy $10^{-5}$ eV. We have constructed stoichiometric slabs of Nb$_3$Sn with Nb$_2$ and NbSn terminations as shown in Fig. \ref{Crystal structure}(c) and the internal degrees of freedom were relaxed.

The density of states (DOS) is a relevant quantity to examine the superconducting properties. We report the DOS Nb$_3$Ge, Nb$_3$Sn and Nb$_3$Sb in Fig. \ref{DOSGe}(a-c). From the DOS, we can see that we have the d-orbitals roughly between -3 eV and +5 eV sandwiched between p-orbitals spectral weight. In these cases, it is extremely challenging to decouple the d-orbitals from the p-orbitals due to their strong hybridization as shown for other materials with such p-d spectral weight\cite{PhysRevMaterials.3.095004,Wadge22}. To perform the wannierization, we must include the d-orbitals of Nb and the p-orbitals of Ge, Sn and Sb. The DOS results were obtained within the VASP code and matched with the results obtained with Quantum Espresso.

We report the calculation of the equilibrium lattice constant within PBEsol and PBE. The system is cubic and does not have internal degrees of freedom, therefore, the only parameter to optimize is the lattice constant $a$. 
The lattice constant results were obtained within the VASP code and matched with the results obtained with Quantum espresso. The PBE exchange-correlation functional overestimates the experimental lattice constant by 1\%, indeed the lattice constant for Nb$_3$Ge, Nb$_3$Sn and Nb$_3$Sb are 5.188 {\AA}, 5.339 {\AA} and 5.314 {\AA}, respectively. The PBEsol gives slightly better results underestimating the lattice constant by 0.1\% (i.e., lattice constant for Nb$_3$Ge, Nb$_3$Sn and Nb$_3$Sb are 5.135 {\AA}, 5.280 {\AA} and 5.259 {\AA}, respectively. 

\section{Phonon dispersions and Migdal-Eliashberg spectral functions for Ta$_3$Y (Y = Ge,Sn,Sb) and Nb$_3$Sb compounds}

In this Section, we report the phonon dispersion curves, phonon density of states, Migdal-Eliashberg spectral functions and the electron-phonon coupling values for all the compounds except that for Nb$_3$Ge and Nb$_3$Sn which are described in the main text. 
The results are reported in Fig. \ref{phonon-ta3y-nb3sb}. The electron-phonon coupling coefficient $\lambda$($\omega$) for Ta$_3$As, Ta$_3$Sb, Ta$_3$Bi and Nb$_3$Sb are 0.56, 0.41, 0.45 and 0.47, respectively. Correspondingly the $T_c$ is slightly higher in Ta$_3$As, Ta$_3$Bi and Nb$_3$Sb as compared to Ta$_3$Sb (see the main text for the numerical values), this is essentially due to the density of states at the Fermi level in the electronic structure. Although, Ta$_3$Sb shows softening of the optical phonon modes which implies that the $T_c$ should be as high as Ta$_3$Bi and Nb$_3$Sb, however, the density of states at the Fermi level in the electronic structure dominates the superconducting behavior leading to lower $T_c$.

\section{Hybridization of the Dirac surface states with the s-bands for Ta-based systems}

In this Section, we want to include the s-band in our tight-binding model and see the effect on the Dirac surface states. While the results of the main text are obtained with high numerical accuracy, the process of including the s-band in the tight-binding Hamiltonian is obtained within strong approximations in the wannierization process (just including the s-band in the frozen window).  
The presence of the s-band in the Ta-based system affects the Dirac surface states by submerging the Dirac points below them. We report in Fig. \ref{with-s-bands-ta3y} the surface states for the Ta$_3$Y which include the s-bands. In our calculations, we can observe the Dirac surface states are pushed below the s-bands and the Dirac surface states are blurred by the hybridization with the s-bands. 

The parity of the s-band and the parity of the other highest valence bands are both positive, therefore, we expect that this s-band would produce an adiabatic transition without affecting the topology\cite{PhysRevB.76.045302,doi:10.1146/annurev-conmatphys-031214-014749}. Indeed, in the literature, it was shown that the Dirac surface states of Ta$_3$Sb can coexist with the s-bands\cite{derunova2019giant}.

\medskip
\newpage
\bibliography{bibliography}
\end{document}